# Stability of equilibrium under constraints: Role of second-order constrained derivatives

Tamás Gál

Department of Theoretical Physics, University of Debrecen, 4010 Debrecen, Hungary
Email address: galt@phys.unideb.hu

**Abstract:** In the stability analysis of an equilibrium, given by a stationary point of a functional $F[\rho]$ (free energy functional, e.g.), the second derivative of $F[\rho]$ plays the essential role. If the system in equilibrium is subject to the conservation constraint of some extensive property (e.g. volume, material, or energy conservation), the Euler equation determining the stationary point corresponding to the equilibrium alters according to the method of Lagrange multipliers. Here, the question as to how the effects of constraints can be taken into account in a stability analysis based on second functional derivatives is examined. It is shown that the concept of constrained second derivatives incorporates all the effects due to constraints; therefore constrained second derivatives provide the proper tool for the stability analysis of equilibria under constraints. For a physically important type of constraints, it is demonstrated how the presented theory works. Further, the rigorous derivation of a recently obtained stability condition for a special case of equilibrium of ultrathin-film binary mixtures is given, presenting a guide for similar analyses.

# I. Introduction

Conservation of some extensive quantity is needed to be accounted for in many physical theories, as in the case of fluid-dynamical models containing conserved order parameters [1], e.g. The dynamics of various systems is often governed by the derivative $\frac{\delta A[\rho]}{\delta \rho(x)}$ of some function(al) $A[\rho]$ of the dynamical variable(s) $\rho(x)$ describing the motion, such as the free-energy functional in fluid dynamics. Since, in general, $A[\rho]$ may have physical relevance only over the domain of $\rho(x)$'s obeying the given conservation constraint(s) $C[\rho] = C$, the equation(s) (of motion) containing $\frac{\delta A[\rho]}{\delta \rho(x)}$ has to be invariant under the replacement of $A[\rho]$ by another functional $A'[\rho]$ that equals $A[\rho]$ for $\rho(x)$'s satisfying the constraint [2]. Therefore, the derivative of $A[\rho]$ has to be modified according to the constraints, leading to the appearance of constrained functional derivatives,

$$\frac{\delta A[\rho]}{\delta_C \rho(x)} = \frac{\delta A[\rho]}{\delta \rho(x)} - \mu[C[\rho]; A[\rho]] \frac{\delta C[\rho]}{\delta \rho(x)} , \qquad (1)$$

introduced in [3]. In many cases (see those in [1], e.g.), this modification of an $\frac{\delta A[\rho]}{\delta \rho(x)}$ is cancelled in the equation(s) of motion for $\rho(x)$ (e.g., due to a $\nabla$ operator); in other words, the form of the equation(s) of motion itself ensures the above-mentioned invariance. However, this cannot be expected in general, especially not for complex constraints – for example, constraints coupling the variables of the given *A*. This is shown also by the fluid-dynamical model proposed by Clarke [4,5] for the description of simultaneous dewetting and phase separation in thin-film binary mixtures [6]. In [5], therefore, the method of constrained differentiation [3] is utilized to set up the equations of motion for this model. (See Ref. [2] for an analysis of that application.)

Beside first(-order) derivatives, second(-order) derivatives also play an essential role in physics – for example, in stationary-point analysis, including the stability analysis of equilibria (see, e.g., [4,7-10] in fluid dynamics), or in nonlinear response theory [11]. Hence the question immediately arises how second derivatives modify under constraints. In this paper, the proper modification of second derivatives will be given, and it will be shown that these constrained second derivatives provide the necessary tool for physics for the analysis of equilibria under conservation constraints. As an example, a physically important type of constraints, of volume and material conservation, will be considered, showing also how the stability condition obtained in [4] for a special case of equilibrium emerges.

Since the appearance of the present work on arXiv, the theory presented in the following has been applied in the stability analysis of droplet growth in supercooled vapors [12] – in the special case of a norm-conserving constraint, accounting for particle number conservation. (The stability condition of Eq.(3) in [12] is a straight consequence of Eq.(74b) below.)

The paper is organized as follows: Sec.II gives the necessary background for the present work, describing the concept of constrained functional derivatives. Further, Sec.II introduces the problem of how to define constrained second derivatives, raising three routes to define them. In Sec.IV, the answer to this question is given; it is shown that one of the definitions proposed in Sec.II incorporates all effects to be taken into account when using second functional derivatives under constraints. Sec.III presents a preliminary to Sec.IV by analyzing a concrete physical example of equilibrium analysis that can be treated in a simplified way – at the same time, it already shows well the effects to account for in a general treatment of constraints.

## II. Constrained second derivatives

### Constrained functional derivatives

In mathematics, the derivative of a functional $A[\rho]$ at some $\rho(x)$ [13] is defined as a functional $D_F(A)[\rho;\Delta\rho]$ that is linear in $\Delta\rho(x)$ and gives

$$D_F(A)[\rho;\Delta\rho] = A[\rho+\Delta\rho] - A[\rho] + o^{(2)}[\rho;\Delta\rho] \qquad (2a)$$

for all $\Delta\rho(x)$'s, where

$$\lim_{\Delta\rho \to 0} \frac{o^{(2)}[\rho;\Delta\rho]}{\|\Delta\rho\|} = 0 \ .$$

This is the so-called Fréchet derivative. Alternatively, the derivative of a functional can be defined as a continuous, linear functional $D_G(A)[\rho;\Delta\rho]$ in $\Delta\rho(x)$ that gives the Gâteaux differential (the directional derivative, in the direction $\Delta\rho(x)$) for any $\Delta\rho(x)$,

$$D_G(A)[\rho;\Delta\rho] = \lim_{\varepsilon \to 0} \frac{A[\rho+\varepsilon\Delta\rho] - A[\rho]}{\varepsilon} \ . \qquad (2b)$$

This derivative is called the Gâteaux derivative. The following theorem throws light upon the connection between the two definitions: If the Fréchet derivative exists at a $\rho(x)$ then the Gâteaux derivative exists there as well and the two derivatives are equal. (This means that the Fréchet definition is a stronger definition.) In physical equations, a functional derivative $D(A)[\rho;\ ]$ appears in a form $\frac{\delta A[\rho]}{\delta\rho(x)}$ defined by

$$\int \frac{\delta A[\rho]}{\delta\rho(x')} \Delta\rho(x')\,dx' = D(A)[\rho;\Delta\rho] \qquad \text{(for all } \Delta\rho(x)\text{'s).} \qquad (3)$$

$\dfrac{\delta A[\rho]}{\delta \rho(x)}$ can be obtained from $D(A)[\rho;\Delta\rho]$ formally by writing $\Delta\rho(x') = \delta(x-x')$; though note that $\delta(x-x')$ can be considered as a function only in a generalized sense – in fact, it is a distribution.

As mentioned in the Introduction, the proper treatment of constraints in a physical theory in general requires the modification of the derivative $\dfrac{\delta A[\rho]}{\delta \rho(x)}$. In [3], the formula

$$\frac{\delta A[\rho]}{\delta_K \rho(x)} = \frac{\delta A[\rho]}{\delta \rho(x)} - \left( \frac{1}{K} \int \frac{f(\rho(x'))}{f^{(1)}(\rho(x'))} \frac{\delta A[\rho]}{\delta \rho(x')} dx' \right) f^{(1)}(\rho(x)) \qquad (4)$$

for that modification has been proposed, for constraints of the form

$$\int f(\rho(x))dx = K \ . \qquad (5)$$

$f$ is an invertible function, which may have an explicit $x$-dependence as well, and $f^{(1)}$ denotes its first derivative. We will call $\dfrac{\delta A[\rho]}{\delta_K \rho(x)}$ a $K$-constrained (or $K$-conserving) derivative. A constrained derivative $\dfrac{\delta A[\rho]}{\delta_K \rho(x)}$ given by Eq.(4) fulfils two essential conditions [14]: (i) two functionals that are equal over a domain of $\rho(x)$'s determined by Eq.(5) (a $K$-restricted domain) should have equal $K$-conserving derivatives over that domain [K-equality condition], and (ii) for a functional $A[\rho]$ that is independent of $K$ of $\rho$ (in the sense that $A[f^{-1}(\lambda f(\rho))] = A[\rho]$, where $\lambda$ is an arbitrary real number),

$$\frac{\delta A[\rho]}{\delta_K \rho(x)} = \frac{\delta A[\rho]}{\delta \rho(x)} \qquad (6)$$

[K-independence condition]. From condition (i),

$$\frac{\delta A[\rho]}{\delta'_K \rho(x)} = \frac{\delta A[\rho]}{\delta \rho(x)} - f^{(1)}(\rho(x)) \int \frac{u(x')}{f^{(1)}(\rho(x'))} \frac{\delta A[\rho]}{\delta \rho(x')} dx' \qquad (7)$$

follows, with $u(x)$ an arbitrary function that integrates to one, while condition (ii) fixes $u(x)$ as $u(x) = \frac{f(\rho(x))}{\int f(\rho(x'))\,dx'}$. For linear constraints (with linear $C[\rho]$), i.e. for constraints $\int g(x)\rho(x)dx = L$, $K$-independence means that $A[\lambda\rho] = A[\rho]$ (for any scalar $\lambda$), i.e., $A[\rho]$ is homogeneous of degree zero, from which

$$\int \rho(x) \frac{\delta A[\rho]}{\delta \rho(x)} dx = 0 \;, \tag{8}$$

giving indeed Eq.(6). In this section in the following, the arguments will be presented considering the special case of the normalization conservation

$$\int \rho(x)\,dx = N \tag{9}$$

for simplicity in presentation; however, their generalization for arbitrary $K$ (and for more complex constraints) is straightforward. For this case, the formula Eq.(4) gives

$$\frac{\delta A[\rho]}{\delta_N \rho(x)} = \frac{\delta A[\rho]}{\delta \rho(x)} - \frac{1}{N} \int \rho(x') \frac{\delta A[\rho]}{\delta \rho(x')} dx' \;. \tag{10}$$

An essential feature of $N$-conserving derivatives is that they deliver the differential, for an arbitrary $\Delta\rho(x)$, that corresponds to the $N$-conserving change of the functional variable, $\Delta_N \rho(x)$; that is,

$$\int \frac{\delta A[\rho]}{\delta_N \rho(x)} \Delta\rho(x)\,dx = \int \frac{\delta A[\rho]}{\delta \rho(x)} \Delta_N \rho(x)\,dx \qquad \text{(for all } \Delta\rho(x)\text{'s)}\;. \tag{11a}$$

An unconstrained $\Delta\rho(x)$ is projected to an $N$-conserving component $\Delta_N \rho(x)$ [$=\tilde{\rho}(x) - \rho(x)$, with $\int \tilde{\rho}(x)dx = \int \rho(x)dx$] via

$$\Delta_N \rho(x) = \int \left\{ \delta(x-x') - \frac{\rho(x)}{N} \right\} \Delta\rho(x')\,dx' = \int \frac{\delta \rho(x)}{\delta_N \rho(x')} \Delta\rho(x')\,dx' \;. \tag{11b}$$

To have Eq.(11a), the projection $\hat{P} = \int dx' \left\{ \delta(x-x') - \frac{\rho(x)}{N} \right\}$ could be replaced by any $\hat{P} = \int dx' \{\delta(x-x') - u(x)\}$, where $u(x)$ integrates to one [14]; i.e., $\frac{\delta \rho(x)}{\delta_N \rho(x')} = \delta(x-x') - u(x)$.

We choose Eq.(11b) in order to have $\frac{\delta A[\rho]}{\delta_N \rho(x)} = \frac{\delta A[\rho]}{\delta \rho(x)}$ for $N$-independent functionals, for which $A[\lambda \rho] = A[\rho]$ [14]. Note that the essential criterion $\hat{P}\hat{P}\Delta\rho(x) = \hat{P}\Delta\rho(x')$ (i.e., $\hat{P}\Delta_N \rho(x') = \Delta_N \rho(x)$) is trivially fulfilled, since $\int \Delta_N \rho(x) dx = 0$.

## How to define constrained second derivatives ?

The second derivative $D^2(A)[\rho; \, , \,]$ of a functional $A[\rho]$ is defined as the derivative of the first derivative $D(A)[\rho; \,]$ of $A[\rho]$ – which is a symmetric bilinear functional. In physics, it is usually represented by $\frac{\delta^2 A[\rho]}{\delta \rho(x) \delta \rho(x')}$ defined by

$$\iint \frac{\delta^2 A[\rho]}{\delta \rho(x) \delta \rho(x')} \Delta\rho(x) \Delta\rho(x') dx dx' = D^2(A)[\rho; \Delta\rho, \Delta\rho] \quad \text{(for all } \Delta\rho(x)\text{'s).} \quad (12)$$

Higher-order derivatives of a functional can be defined similarly. (Note that higher Gâteaux derivatives will give the higher Gâteaux differentials, or variations, $D_G^n(A)[\rho;(\Delta\rho)^n] = \left. \frac{d^n}{dt^n} A[\rho + t\Delta\rho] \right|_{t=0}$, in any direction $\Delta\rho(x)$.) With higher derivatives $\frac{\delta^n A}{\delta \rho^n}$, then, a functional can be given by a Taylor expansion, with remainder,

$$A[\rho + \Delta\rho] = A[\rho] + \int \frac{\delta A[\rho]}{\delta \rho(x)} \Delta\rho(x) dx + \frac{1}{2!} \iint \frac{\delta^2 A[\rho]}{\delta \rho(x) \delta \rho(x')} \Delta\rho(x) \Delta\rho(x') dx dx' + \dots . \quad (13)$$

It is worth noting that an advantage of the Gâteaux concept of a functional derivative is that the Gâteaux differential (i.e., the directional derivative) is defined without the Gâteaux derivative (and, e.g., a Taylor-like expansion along a $\Delta\rho(x)$ direction may be given by the $n$th Gâteaux differentials, $n=0,1,\ldots$, without existing $n$th Gâteaux derivatives) – while the Fréchet differential and the Fréchet derivative are attached concepts.

Taking the $N$-conserving derivative of the first $N$-conserving derivative of a functional $A[\rho]$ yields

$$\frac{\delta}{\delta_N\rho(x)}\frac{\delta A[\rho]}{\delta_N\rho(x')} = \frac{\delta^2 A[\rho]}{\delta\rho(x)\delta\rho(x')} - \frac{1}{N}\frac{\delta A[\rho]}{\delta_N\rho(x)} - \frac{1}{N}\left(\int\rho(x'')\frac{\delta^2 A[\rho]}{\delta\rho(x'')\delta\rho(x')}dx'' + \int\rho(x'')\frac{\delta^2 A[\rho]}{\delta\rho(x)\delta\rho(x'')}dx''\right)$$

$$+\frac{1}{N^2}\iint\rho(x'')\rho(x''')\frac{\delta^2 A[\rho]}{\delta\rho(x'')\delta\rho(x''')}dx''dx''' \ . \tag{14}$$

Eq.(14) transforms a symmetric $\frac{\delta^2 A[\rho]}{\delta\rho(x)\delta\rho(x')}$ (in $x$ and $x'$) into an asymmetric second derivative [3], due to a single term $-\frac{1}{N}\frac{\delta A[\rho]}{\delta_N\rho(x)}$. However, although second derivatives $D^2(A)$ emerge as the derivatives of first derivatives, it is not necessarily the proper way of defining $\frac{\delta^2 A}{\delta_N\rho^2}$. (Note that even for $\frac{\delta^2 A}{\delta\rho^2}$, defined by Eq.(12), $\frac{\delta^2 A}{\delta\rho^2} = \frac{\delta}{\delta\rho}\frac{\delta A}{\delta\rho}$ has to be justified.)

In [15], another possible way to define constrained second (and higher) derivatives is given. Following the idea of Eq.(11),

$$\iint\frac{\delta^2 A[\rho]}{\delta_N\rho(x)\delta_N\rho(x')}\Delta\rho(x)\Delta\rho(x')dxdx' = \iint\frac{\delta^2 A[\rho]}{\delta\rho(x)\delta\rho(x')}\Delta_N\rho(x)\Delta_N\rho(x')dxdx' \ \text{(for all } \Delta\rho(x)\text{'s)},$$

$$\tag{15}$$

arises, which yields the formula

$$\frac{\delta^2 A[\rho]}{\delta_N\rho(x)\delta_N\rho(x')} = \frac{\delta^2 A[\rho]}{\delta\rho(x)\delta\rho(x')} - \frac{1}{N}\int\rho(x'')\left(\frac{\delta^2 A[\rho]}{\delta\rho(x)\delta\rho(x'')} + \frac{\delta^2 A[\rho]}{\delta\rho(x'')\delta\rho(x')}\right)dx''$$

$$+\frac{1}{N^2}\iint \rho(x'')\rho(x''')\frac{\delta^2 A[\rho]}{\delta\rho(x'')\delta\rho(x''')}dx''dx''' \left(=\frac{\delta}{\delta_N\rho(x)}\frac{\delta A[\rho]}{\delta_N\rho(x')}+\frac{1}{N}\frac{\delta A[\rho]}{\delta_N\rho(x)}\right). \quad (16)$$

This $\dfrac{\delta^2 A}{\delta_N\rho^2}$ satisfies the K-equality condition for second derivatives, since the second derivatives of two functionals that are equal on a domain of $\rho_N(x)$'s may differ only by some $\mu(x)+\mu'(x')$ on that domain, which difference is cancelled in Eq.(16). On the other hand, $\dfrac{\delta^2 A}{\delta_N\rho^2}$ of Eq.(16) does not satisfy the K-independence condition for second derivatives. For a degree-zero homogeneous $A[\rho]$,

$$\int \rho(x')\left(\frac{\delta^2 A[\rho]}{\delta\rho(x)\delta\rho(x')}\right)dx' = -\frac{\delta A[\rho]}{\delta\rho(x)} \quad (17)$$

– which comes from differentiating Eq.(8) with respect to $\rho(x)$. It can be seen that Eq.(16) does not become $\dfrac{\delta^2 A}{\delta\rho^2}$ for degree-zero homogeneous $A[\rho]$s: the terms beside $\dfrac{\delta^2 A}{\delta\rho^2}$ in Eq.(16) do not vanish. [A false logic could make one say that the K-independence condition is satisfied by Eq.(16): If $A[\rho]$ is degree-one homogeneous, hence

$$\int \rho(x')\left(\frac{\delta^2 A[\rho]}{\delta\rho(x)\delta\rho(x')}\right)dx' = 0$$

holds for its second derivative (following from $\int \rho(x)\dfrac{\delta A[\rho]}{\delta\rho(x)}dx = A[\rho]$), then $\dfrac{\delta A[\rho]}{\delta_N\rho(x)}$ is degree-zero homogeneous; *consequently*, $A[\rho]$'s $N$-conserving second derivative should reduce to $\dfrac{\delta^2 A[\rho]}{\delta\rho(x)\delta\rho(x')}$, which it indeed does, according to the relation above. However, in this case, it is of course $\dfrac{\delta}{\delta_N\rho(x)}\dfrac{\delta A[\rho]}{\delta_N\rho(x')}$ what should give back the "unconstrained second" derivative – which it does, giving back $\dfrac{\delta}{\delta\rho(x)}\dfrac{\delta A[\rho]}{\delta_N\rho(x')}$.]

Directly following the route [14] that defines the *N*-conserving derivative of a functional $A[\rho]$ (over a domain of $\rho(x)$'s of a given *N*) as the unconstrained derivative of $A[\rho_N]$'s degree-zero homogeneous extension,

$$\frac{\delta A[\rho]}{\delta_N \rho(x)} := \left.\frac{\delta A[\rho_N^0[\rho]]}{\delta \rho(x)}\right|_{\int \rho = N}, \qquad (18)$$

with $\rho_N^0[\rho] = \dfrac{N}{\int \rho(x')dx'}\rho(x)$, however, yields a definition of $\dfrac{\delta^2 A[\rho]}{\delta_N \rho(x) \delta_N \rho(x')}$ that satisfies also the K-independence condition. Defining an *N*-constrained second derivative as

$$\frac{\delta^2 A[\rho]}{\delta_N \rho(x) \delta_N \rho(x')} := \left.\frac{\delta^2 A[\rho_N^0[\rho]]}{\delta \rho(x) \delta \rho(x')}\right|_{\int \rho = N} \qquad (19)$$

leads to

$$\frac{\delta^2 A[\rho]}{\delta_N \rho(x)\delta_N \rho(x')} = \frac{\delta^2 A[\rho]}{\delta\rho(x)\delta\rho(x')} - \frac{1}{N}\left(\frac{\delta A[\rho]}{\delta_N \rho(x)} + \frac{\delta A[\rho]}{\delta_N \rho(x')}\right)$$

$$-\frac{1}{N}\int \rho(x'')\left(\frac{\delta^2 A[\rho]}{\delta\rho(x)\delta\rho(x'')} + \frac{\delta^2 A[\rho]}{\delta\rho(x'')\delta\rho(x')}\right)dx'' + \frac{1}{N^2}\iint \rho(x'')\rho(x''')\frac{\delta^2 A[\rho]}{\delta\rho(x'')\delta\rho(x''')}dx''dx'''$$

$$\left(= \frac{\delta}{\delta_N \rho(x)}\frac{\delta A[\rho]}{\delta_N \rho(x')} - \frac{1}{N}\frac{\delta A[\rho]}{\delta_N \rho(x')}\right). \qquad (20)$$

Eq.(20) indeed fulfils the K-independence condition (and also the K-equality condition) of second derivatives – which can be seen by using Eq.(17). Notice that, similarly to Eq.(16), Eq.(20) preserves the symmetry of a symmetric $\dfrac{\delta^2 A[\rho]}{\delta \rho(x)\delta\rho(x')}$ (in *x* and *x'*); however, instead of simply cancelling the "undesirable" term, Eq.(20) symmetrizes it.

Having the three above possible definitions, Eqs.(14), (16) and (20), of an *N*-conserving second derivative, the question is which one we should choose. (Note that the three formulae yields the same result at a stationary $\rho(x)$ under Eq.(9), where $\dfrac{\delta A[\rho]}{\delta_N \rho(x)}=0$, but for a non-stationary $\rho(x)$, and for other constraints, this is not so in general.) Shall it be Eq.(14), which may be the straightest in origination, but there is no conceptual reason behind it? Or Eq.(20), which fulfils the K-independence condition? Or Eq.(16), not containing the first derivative of $A[\rho]$, which undoubtedly seems to be a reasonable feature? This question will be answered in the following through considering the concrete case of a physically important complex constraint, taken from the thin liquid film model [4] proposed by Clarke (the dynamics of which was given later [5] by the application of constrained differentiation). To gain insight into the effects of constraints that have to be taken into account in a stability analysis of equilibrium, first, in the next section, we will re-consider Clarke's treatment [4] of a special case of equilibrium in his model, and clarify its mathematical aspects.

## III. A case of equilibrium under constraint, coupling the variables of the free-energy functional

In [4], Clarke examines the stability of equilibrium of a model of thin-film binary mixtures, described by two variables, the height $h(x)$ and the composition $\phi(x)$. The examined equilibrium, with $h(x) = h_0$ and $\phi(x) = \phi_0$ (flat, homogeneous distribution), corresponds to a stationary point of the free-energy functional $F_T[h,\phi]$, built from $h(x)$ and $\phi(x)$, under the constraint of volume and material conservation,

$$\int h(x)\,dx = N \tag{21}$$

and

$$\int \phi(x)h(x)\,dx = B \ , \tag{22}$$

respectively (with $N = A h_0$, and $B = A\phi_0 h_0$, $A$ denoting the area of the film). The emerging Euler-Lagrange equations are

$$\frac{\delta F_T[h,\phi]}{\delta h(x)} = \mu_1 + \mu_2 \phi(x) \tag{23a}$$

and

$$\frac{\delta F_T[h,\phi]}{\delta \phi(x)} = \mu_2 h(x), \tag{23b}$$

which, utilizing that $F_T[h,\phi]$ is taken in a form $F_T[h,\phi] = \int f_T(h(x),\phi(x))\,dx$, give

$$\frac{\partial f_T(h_0,\phi_0)}{\partial h} = \mu_1 + \mu_2 \phi_0 \tag{24a}$$

and

$$\frac{\partial f_T(h_0,\phi_0)}{\partial \phi} = \mu_2 h_0 \tag{24b}$$

for the constant $(h_0,\phi_0)$.

To obtain the condition of instability of the above state, the problem is traced back in [4] to the stability analysis of simple, two-variable functions $g(y_1,y_2)$. In that case, the sufficient condition for a local maximum at a given stationary point (i.e., an instable stationary point in the case of the free energy) is well-known to be

$$\frac{\partial^2 g(y_1,y_2)}{\partial y_1^2}\frac{\partial^2 g(y_1,y_2)}{\partial y_2^2} - \left(\frac{\partial^2 g(y_1,y_2)}{\partial y_1 \partial y_2}\right)^2 < 0. \tag{25}$$

(The expression on the left side in Eq.(25) is the determinant of the Hessian matrix of $g(y_1,y_2)$.) To account for the constraints (21) and (22), Clarke makes use of

$$\int \Delta h(x)\,dx = 0 \tag{26}$$

and

$$h_0 \int \Delta \phi(x)\,dx + \phi_0 \int \Delta h(x)\,dx = -\int \Delta \phi(x) \Delta h(x)\,dx \tag{27}$$

(coming from Eqs.(21) and (22), respectively) in the Taylor expansion of the free-energy,

$$F_T[h+\Delta h,\phi+\Delta\phi] = F_T[h,\phi] + \int \frac{\delta F_T[h,\phi]}{\delta h(x)}\Delta h(x)\,dx + \int \frac{\delta F_T[h,\phi]}{\delta\phi(x)}\Delta\phi(x)\,dx$$

$$+\frac{1}{2}\iint \frac{\delta^2 F_T[h,\phi]}{\delta h(x)\delta h(x')}\Delta h(x)\Delta h(x')\,dxdx' + \frac{1}{2}\iint \frac{\delta^2 F_T[h,\phi]}{\delta\phi(x)\delta h(x')}\Delta\phi(x)\Delta h(x')\,dxdx'$$

$$+\frac{1}{2}\iint \frac{\delta^2 F_T[h,\phi]}{\delta h(x)\delta\phi(x')}\Delta h(x)\Delta\phi(x')\,dxdx' + \frac{1}{2}\iint \frac{\delta^2 F_T[h,\phi]}{\delta\phi(x)\delta\phi(x')}\Delta\phi(x)\Delta\phi(x')\,dxdx'$$

$$+ \text{ higher-order terms}, \tag{28}$$

taken at $(h_0,\phi_0)$,

$$F_T[h_0+\Delta h,\phi_0+\Delta\phi] = F_T[h_0,\phi_0] + \frac{\partial f_T(h_0,\phi_0)}{\partial h}\int \Delta h(x)\,dx + \frac{\partial f_T(h_0,\phi_0)}{\partial\phi}\int \Delta\phi(x)\,dx$$

$$+\frac{1}{2}\frac{\partial^2 f_T(h_0,\phi_0)}{\partial h^2}\int (\Delta h(x))^2\,dx + \frac{\partial^2 f_T(h_0,\phi_0)}{\partial h\,\partial\phi}\int \Delta\phi(x)\Delta h(x)\,dx$$

$$+\frac{1}{2}\frac{\partial^2 f_T(h_0,\phi_0)}{\partial\phi^2}\int (\Delta\phi(x))^2\,dx + \ldots, \tag{29}$$

to obtain

$$F_T[h_0+\Delta h,\phi_0+\Delta\phi] = F_T[h_0,\phi_0] + \frac{1}{2}\frac{\partial^2 f_T(h_0,\phi_0)}{\partial h^2}\int (\Delta h(x))^2\,dx$$

$$+\left(\frac{\partial^2 f_T(h_0,\phi_0)}{\partial h\,\partial\phi} - \mu_2\right)\int \Delta\phi(x)\Delta h(x)\,dx + \frac{1}{2}\frac{\partial^2 f_T(h_0,\phi_0)}{\partial\phi^2}\int (\Delta\phi(x))^2\,dx + \ldots$$

$$\tag{30}$$

The variations $\Delta h$ and $\Delta\phi$ are then handled by their Fourier series expansions,

$$\Delta h(x) = \frac{\Delta h_0}{2} + \Delta h_{1a}\sin\frac{2\pi x}{A} + \Delta h_{1b}\cos\frac{2\pi x}{A} + \ldots \tag{31a}$$

and

$$\Delta\phi(x) = \frac{\Delta\phi_0}{2} + \Delta\phi_{1a}\sin\frac{2\pi x}{A} + \Delta\phi_{1b}\cos\frac{2\pi x}{A} + \ldots, \tag{31b}$$

respectively, where

$$\Delta h_0 \equiv \frac{2}{A} \int \Delta h(x) \, dx = 0 \;, \tag{32}$$

due to Eq.(26). (For simplicity, $x$ is taken to be of one dimension, instead of the two dimensions in [4].) The integrals in Eq.(30) thus become

$$\int (\Delta h(x))^2 \, dx = \frac{A}{2} \left( \frac{1}{2} (\Delta h_0)^2 + (\Delta h_{1a})^2 + (\Delta h_{1b})^2 + \ldots \right) = \frac{A}{2} \left( \frac{\Delta h_0}{\sqrt{2}}, \Delta h_{1a}, \Delta h_{1b}, \ldots \right)^2 , \tag{33}$$

$$\int \Delta \phi(x) \Delta h(x) \, dx = \frac{A}{2} \left( \frac{1}{2} \Delta \phi_0 \Delta h_0 + \Delta \phi_{1a} \Delta h_{1a} + \ldots \right) = \frac{A}{2} \left( \frac{\Delta \phi_0}{\sqrt{2}}, \Delta \phi_{1a}, \ldots \right) \left( \frac{\Delta h_0}{\sqrt{2}}, \Delta h_{1a}, \ldots \right) , \tag{34}$$

and

$$\int (\Delta \phi(x))^2 \, dx = \frac{A}{2} \left( \frac{1}{2} (\Delta \phi_0)^2 + (\Delta \phi_{1a})^2 + \ldots \right) = \frac{A}{2} \left( \frac{\Delta \phi_0}{\sqrt{2}}, \Delta \phi_{1a}, \ldots \right)^2 ; \tag{35}$$

consequently,

$$F_T[h_0 + \Delta h, \phi_0 + \Delta \phi] = F_T[h_0, \phi_0] + \frac{1}{2} \frac{\partial^2 f_T(h_0, \phi_0)}{\partial h^2} \frac{A}{2} \left( \frac{\Delta h_0}{\sqrt{2}}, \Delta h_{1a}, \ldots \right)^2$$

$$+ \left( \frac{\partial^2 f_T(h_0, \phi_0)}{\partial h \, \partial \phi} - \mu_2 \right) \frac{A}{2} \left( \frac{\Delta \phi_0}{\sqrt{2}}, \Delta \phi_{1a}, \ldots \right) \left( \frac{\Delta h_0}{\sqrt{2}}, \Delta h_{1a}, \ldots \right)$$

$$+ \frac{1}{2} \frac{\partial^2 f_T(h_0, \phi_0)}{\partial \phi^2} \frac{A}{2} \left( \frac{\Delta \phi_0}{\sqrt{2}}, \Delta \phi_{1a}, \ldots \right)^2 + \ldots . \tag{36}$$

Finally, an account for the constraints, as regards the variations, i.e. Eq.(32) and

$$-h_0 \Delta \phi_0 = \Delta \phi_{1a} \Delta h_{1a} + \Delta \phi_{1b} \Delta h_{1b} + \ldots \tag{37}$$

[coming from Eq.(27), with Eq.(34), and $\frac{2}{A} \int \Delta \phi(x) \, dx = \Delta \phi_0$, and Eq.(32)], has to be made. Eq.(37) can be inserted into Eq.(35) [and into Eq.(34) in a case Eq.(21) is not required], in the place of $\Delta \phi_0$. The obtained term being of order higher than two, this yields

$$F_T[h_0 + \Delta h, \phi_0 + \Delta \phi] = F_T[h_0, \phi_0] + \frac{1}{2} \frac{\partial^2 f_T(h_0, \phi_0)}{\partial h^2} \frac{A}{2} (\Delta h_{1a}, \ldots)^2$$

$$+\left(\frac{\partial^2 f_T(h_0,\phi_0)}{\partial h\,\partial\phi}-\mu_2\right)\frac{A}{2}(\Delta\phi_{1a},\ldots)(\Delta h_{1a},\ldots)$$

$$+\frac{1}{2}\frac{\partial^2 f_T(h_0,\phi_0)}{\partial\phi^2}\frac{A}{2}(\Delta\phi_{1a},\ldots)^2 + \text{higher-order terms}. \tag{38}$$

Since $(\Delta h_{1a},\ldots)$ and $(\Delta\phi_{1a},\ldots)$ can now be varied freely, by fixing all but a pair of $\Delta h_i$ and $\Delta\phi_j$ as 0, Eq.(38) gives a two-variable (second-order) variational problem, yielding

$$\frac{\partial^2 f_T(h_0,\phi_0)}{\partial h^2}\frac{\partial^2 f_T(h_0,\phi_0)}{\partial\phi^2}-\left(\frac{\partial^2 f_T(h_0,\phi_0)}{\partial h\,\partial\phi}-\frac{1}{h_0}\frac{\partial f_T(h_0,\phi_0)}{\partial\phi}\right)^2 < 0 \tag{39}$$

as the criterion of instability, on the basis of Eq.(25). (Clarke has obtained this result in a somewhat more complicated fashion actually, the mathematical essence of which, however, is just that described above. Note also that in [4], $f_T$ is divided into two parts, $f_T(h,\phi)=h f_b(\phi)+f_s(h,\phi)$, which has no relevance here.)

The stability condition Eq.(39) has been verified also through the equations of motion of $(h(x),\phi(x))$ set up in [5]. However, two essential questions arise with respect to the above derivation in the context of more general situations. First, why should Eq.(27) (with Eq.(26)) be utilized in the way utilized above, obtaining Eq.(30), cancelling the $\int\Delta\phi(x)\,dx$ term completely? Second, does not the occurrence of the new higher-order terms, due to the use of Eq.(37), matter at all with respect to the stability analysis? (Note that this actually leads to the interesting fact that formally even a free variation of $(2^{-1/2}\Delta h_0,\Delta h_{1a},\ldots)$ and $(2^{-1/2}\Delta\phi_0,\Delta\phi_{1a},\ldots)$ in Eq.(36) yields the result Eq.(39); though of course the constraint Eq.(27) is already taken into account in Eq.(36) in some way.)

The answers to these questions are reassuring. For, in the proof of the theorem (from which the criterion Eq.(25) in the case of two-variable ordinary functions also follows) that the second derivative $D^2(A)[\rho;\ ,\ ]$ of a functional $A[\rho]$ is nonnegative for all $\Delta\rho(x)$'s at a local minimum of $A[\rho]$ (and reverse for a maximum) [16,13], i.e.

$$\iint \frac{\delta^2 A[\rho]}{\delta\rho(x)\delta\rho(x')}\Delta\rho(x)\Delta\rho(x')dxdx' \geq (\leq) 0 \qquad \text{for all } \Delta\rho(x)\text{'s}, \qquad (40)$$

two essential elements are that (i) the first derivative $D(A)[\rho;\ ]$ vanishes, and (ii) higher-than-second order terms become zero. [With strict inequalities, more precisely, with $\geq p > 0$ ($\leq p < 0$), in Eq.(40), and with some restriction on $\Delta\rho(x)$'s (see below Eq.(74b), the condition becomes a sufficient condition for a local minimum (maximum).] For more complex situations, especially in the case of nonlinear constraints (like Eqs.(21) and (22), too), however, another important issue should be accounted for.

Eq.(40) has already been generalized for the case of constraints present, based on Ljusternik's theorems [13]. It can be written as

$$D^2(A)[\rho;\Delta_{\bar{C}}\rho,\Delta_{\bar{C}}\rho] - \mu D^2(C)[\rho;\Delta_{\bar{C}}\rho,\Delta_{\bar{C}}\rho] \geq (\leq) 0 \qquad \text{for all } \Delta_{\bar{C}}\rho(x)\text{'s}, \qquad (41a)$$

i.e.,

$$\iint \left(\frac{\delta^2 A[\rho]}{\delta\rho(x)\delta\rho(x')} - \mu\frac{\delta^2 C[\rho]}{\delta\rho(x)\delta\rho(x')}\right)\Delta_{\bar{C}}\rho(x)\Delta_{\bar{C}}\rho(x')dxdx' \geq (\leq) 0 \qquad \text{for all } \Delta_{\bar{C}}\rho(x)\text{'s} \quad (41b)$$

(with strict inequalities in the "sufficient" version; see note below Eq.(74b)). $\mu$ is the Lagrange multiplier corresponding to the constraint $C[\rho]=C$, which is given by $\mu = \dfrac{\delta A[\rho]}{\delta\rho(x)} \bigg/ \dfrac{\delta C[\rho]}{\delta\rho(x)}$ at the stationary $\rho$. $\Delta_{\bar{C}}\rho(x)$ is such that

$$\int \frac{\delta C[\rho]}{\delta\rho(x)}\Delta_{\bar{C}}\rho(x) = 0 \ ; \qquad (42)$$

that is, $\Delta_{\bar{C}}\rho(x)$ does not have to satisfy the constraint itself but the first derivative of the constraint – which is the same only for linear $C[\rho]$s. A difficulty with the use of Eq.(41) is that the variations $\Delta_{\bar{C}}\rho(x)$ are not free; the fulfillment of Eq.(42) has to be ensured somehow. As will be shown in the next section, this problem can be solved by the use of constrained second derivatives in the place of unconstrained second derivatives, by which the constraint on the variations $\Delta_{\bar{C}}\rho(x)$ can be eliminated. Further, constrained second derivatives incorporate the Lagrange multiplier in Eq.(41), too, providing a natural general treatment of constraints in the stability analysis of equilibria.

## IV. The use of constrained second derivatives in stability analysis of equilibrium with constraints

Taking the constraints (21) and (22) into account in the variations $\Delta h(x)$ and $\Delta\phi(x)$ in Eq.(28) was possible to be done in the way done in Sec.III because of the constancy of the first derivatives in $x$, and the form $const.\cdot\delta(x-x')$ of the second derivatives (second derivative kernels). These simplifying circumstances came from the simple nature of the considered equilibrium [namely, $h_0(x)$ and $\phi_0(x)$ are constant], and that $f_T$ of $F_T[h,\phi] = \int f_T(x)\,dx$ in the considered model depends on $x$ only via the functional variables $h(x)$ and $\phi(x)$. In this section, we wish to treat the general situation. We will consider a general constraint

$$C[\rho] = C , \qquad (43)$$

which may be multi-component, and $\rho(x)$ may denote many variables.

### A. Relaxing the constraint on varying the functional variable by the use of constrained derivatives

Consider the Taylor expansion of the functional $A[\rho]$ above a domain determined by some constraint Eq.(43),

$$A[\rho+\Delta_C\rho] = A[\rho] + \int \frac{\delta A[\rho]}{\delta\rho(x)}\Delta_C\rho(x)dx + \frac{1}{2!}\iint \frac{\delta^2 A[\rho]}{\delta\rho(x)\delta\rho(x')}\Delta_C\rho(x)\Delta_C\rho(x')dxdx'$$

$$+ \frac{1}{3!}\iiint \frac{\delta^3 A[\rho]}{\delta\rho(x)\delta\rho(x')\delta\rho(x'')}\Delta_C\rho(x)\Delta_C\rho(x')\Delta_C\rho(x'')dxdx'dx'' + ... \quad (44)$$

In the case $\rho(x)$ is an $n$-component variable, the second derivative $\frac{\delta^2 A[\rho]}{\delta\rho(x)\delta\rho(x')}$ above will be an $n\times n$ matrix, the third derivative an $n\times n\times n$ matrix, etc. $\Delta_C\rho(x)$ satisfies the constraint, that is,

$$C[\rho+\Delta_C\rho] - C[\rho] = 0 . \quad (45)$$

By expanding $C[\rho+\Delta_C\rho]$ into its Taylor expansion, this gives

$$\int \frac{\delta C[\rho]}{\delta\rho(x)}\Delta_C\rho(x)dx + \frac{1}{2!}\iint \frac{\delta^2 C[\rho]}{\delta\rho(x)\delta\rho(x')}\Delta_C\rho(x)\Delta_C\rho(x')dxdx'$$

$$+ \frac{1}{3!}\iiint \frac{\delta^3 C[\rho]}{\delta\rho(x)\delta\rho(x')\delta\rho(x'')}\Delta_C\rho(x)\Delta_C\rho(x')\Delta_C\rho(x'')dxdx'dx'' + ... = 0 . \quad (46)$$

If $\rho(x)$ is a stationary point of $A[\rho]$ under the constraint Eq.(43), it satisfies the Euler-Lagrange equation

$$\frac{\delta A[\rho]}{\delta\rho(x)} = \mu \frac{\delta C[\rho]}{\delta\rho(x)} , \quad (47)$$

where the Lagrange multiplier $\mu$ is constant with respect to $x$. With the use of Eqs.(46) and (47), the first-order term in Eq.(44) can be eliminated, obtaining

$$A[\rho+\Delta_C\rho] = A[\rho] + \frac{1}{2!}\iint \left(\frac{\delta^2 A[\rho]}{\delta\rho(x)\delta\rho(x')} - \mu\frac{\delta^2 C[\rho]}{\delta\rho(x)\delta\rho(x')}\right)\Delta_C\rho(x)\Delta_C\rho(x')dxdx'$$

$$+ \frac{1}{3!}\iiint \left(\frac{\delta^3 A[\rho]}{\delta\rho(x)\delta\rho(x')\delta\rho(x'')} - \mu\frac{\delta^3 C[\rho]}{\delta\rho(x)\delta\rho(x')\delta\rho(x'')}\right)\Delta_C\rho(x)\Delta_C\rho(x')\Delta_C\rho(x'')dxdx'dx'' + ... \quad (48)$$

In order to free the variations $\Delta_C\rho(x)$ from the constraint, we introduce a mapping $\rho_C[\rho]$ with the following properties: (i) $\rho_C[\rho]$ maps any $\rho(x)$ onto a $\rho_C(x)$, which satisfies the constraint Eq.(43), and (ii) $\rho_C[\rho]$ becomes an identity for $\rho_C(x)$'s, i.e., $\rho_C[\tilde{\rho}_C(x')] = \tilde{\rho}_C(x)$. For an arbitrary change of $\rho(x)$ with a given $C$, this $\rho_C[\rho]$ then gives a $C$-conserving change of $\rho(x)$ via $\Delta_C \rho(x) = \rho_C[\rho + \Delta\rho] - \rho_C[\rho]$. This yields an expansion of $\Delta_C \rho(x)$ in terms of unconstrained variations,

$$\Delta_C \rho(x) = \int \frac{\delta \rho_C[\rho](x)}{\delta \rho(x')} \Delta\rho(x') dx' + \frac{1}{2} \iint \frac{\delta^2 \rho_C[\rho](x)}{\delta \rho(x') \delta \rho(x'')} \Delta\rho(x') \Delta\rho(x'') dx' dx'' + \ldots \quad (49)$$

With the use of the definition

$$\frac{\delta^m A[\rho]}{\delta'_C \rho^m} := \left. \frac{\delta^m A[\rho_C[\rho]]}{\delta \rho^m} \right|_{\rho = \rho_C} \quad (50)$$

(following Eqs.(18) and (19)), Eq.(49) can be written as

$$\Delta_C \rho(x) = \int \frac{\delta \rho(x)}{\delta'_C \rho(x')} \Delta\rho(x') dx' + \frac{1}{2} \iint \frac{\delta^2 \rho(x)}{\delta'_C \rho(x') \delta'_C \rho(x'')} \Delta\rho(x') \Delta\rho(x'') dx' dx'' + \ldots \quad (51)$$

The prime of $\delta'_C$ denotes that the $C$-constrained derivative of Eq.(50) is not required to fulfill a condition like the K-independence condition, in addition to the C-equality condition; that is, $\rho_C[\rho]$ does not have to be "homogeneous" [14] of degree zero. This means that there is some freedom in choosing $\rho_C[\rho]$ to obtain a constrained derivative via Eq.(50). In the following, however, we will write $\delta_C$ instead of $\delta'_C$ for simplicity – but $\delta_K$ will still stand for a $K$-constrained derivative defined via a degree-zero (K-)homogeneous $\rho_K[\rho]$ (i.e., $\rho_K^0[\rho]$).

Insertion of Eq.(51) into Eq.(48) gives

$$A[\rho + \Delta_C \rho] = A[\rho] + \frac{1}{2!} \iint \left( \left( \frac{\delta^2 A[\rho]}{\delta_C \rho(x) \delta_C \rho(x')} \right)^* - \mu \left( \frac{\delta^2 C[\rho]}{\delta_C \rho(x) \delta_C \rho(x')} \right)^* \right) \Delta\rho(x) \Delta\rho(x') dx dx'$$

$$+ \text{ higher-order terms in } \Delta\rho(x), \quad (52)$$

with

$$\left( \frac{\delta^2 A[\rho]}{\delta_C \rho(x) \delta_C \rho(x')} \right)^* = \iint \frac{\delta^2 A[\rho]}{\delta \rho(x'') \delta \rho(x''')} \frac{\delta \rho(x'')}{\delta_C \rho(x)} \frac{\delta \rho(x''')}{\delta_C \rho(x')} dx'' dx''' . \quad (53)$$

Eq.(53) is nothing else than the constrained second derivative defined according to Eq.(11). The * is to distinguish this definition from the one given by Eq.(50). Eq.(52) gives us

$$\iint\left[\left(\frac{\delta^2 A[\rho]}{\delta_C\rho(x)\delta_C\rho(x')}\right)^* - \mu\left(\frac{\delta^2 C[\rho]}{\delta_C\rho(x)\delta_C\rho(x')}\right)^*\right]\Delta\rho(x)\Delta\rho(x')dxdx' \geq (\leq) 0 \quad \text{for all } \Delta\rho(x) \quad (54)$$

as the necessary condition for a local minimum (maximum) in the place of Eq.(41). (Similar to Eq.(41), this becomes a sufficient condition if we replace $\geq/\leq$ with $\geq p >/\leq p <$.) This is justified by the fact that $\Delta_{\bar{C}}\rho(x)$ in Eq.(41) can be written with the help of $\rho_C[\rho]$ as

$$\Delta_{\bar{C}}\rho(x) = \int \frac{\delta\rho(x)}{\delta_C\rho(x')}\Delta\rho(x')dx' \;, \quad (55)$$

and the second-order term in Eq.(52) has been obtained via the first term of Eq.(51). To prove that Eq.(55) is indeed a variation satisfying Eq.(42), just insert it into Eq.(42), and use

$$\frac{\delta A[\rho]}{\delta_C\rho(x)} = \int \frac{\delta A[\rho]}{\delta\rho(x')}\frac{\delta\rho(x')}{\delta_C\rho(x')}dx' \;, \quad (56)$$

and

$$\frac{\delta C[\rho]}{\delta_C\rho(x)} = 0 \;. \quad (57)$$

(The latter equation is a trivial generalization of $\dfrac{\delta K[\rho]}{\delta_K\rho(x)}=0$ [3,14] – a straight consequence of the K-equality condition.)

To illustrate the above through a physical example, we will now give the corresponding expressions for Clarke's model, i.e., for a two-variable functional $F_T[h,\phi]$ with the constraint of Eqs.(21) and (22). This constraint has a general enough form to have a general relevance, and be worth giving for it the corresponding constrained derivatives. It is also relatively simple, but complex enough to throw light onto the character of constrained second derivatives.

The variational form of Eq.(22), corresponding to Eqs.(44) and (46), is

$$\int \Delta\phi(x)h(x)\,dx + \int \phi(x)\Delta h(x)\,dx = -\int \Delta\phi(x)\Delta h(x)\,dx\,, \tag{58a}$$

or

$$\int \Delta\phi(x)h(x)\,dx + \int \phi(x)\Delta h(x)\,dx = -\iint \delta(x-x')\Delta\phi(x)\Delta h(x')\,dxdx'\,. \tag{58b}$$

The corresponding expression for Eq.(21) has already been given; see Eq.(26). With the use of the Euler-Lagrange equations (23), and Eqs.(26) and (58), the Taylor expansion of $F_T[h,\phi]$ [Eq.(28)] can be written over the constrained domain as

$$F_T[h+\Delta_K h,\phi+\Delta_K\phi] = F_T[h,\phi] + \frac{1}{2}\iint \frac{\delta^2 F_T[h,\phi]}{\delta h(x)\delta h(x')}\Delta_K h(x)\Delta_K h(x')\,dxdx'$$

$$+ \iint \left(\frac{\delta^2 F_T[h,\phi]}{\delta\phi(x)\delta h(x')} - \mu_2\delta(x-x')\right)\Delta_K\phi(x)\Delta_K h(x')\,dxdx'$$

$$+ \frac{1}{2}\iint \frac{\delta^2 F_T[h,\phi]}{\delta\phi(x)\delta\phi(x')}\Delta_K\phi(x)\Delta_K\phi(x')\,dxdx' + \text{higher-order terms}\,. \tag{59}$$

The index $K$ denotes that the variations obey Eqs.(26) and (58) – this notation was not used in Eqs.(28)-(30) and (36) for simplicity.

The expansions for $\Delta_K h(x)$ and $\Delta_K\phi(x)$ to be inserted into Eq.(59) can be given as

$$\Delta_K h(x) = \int \frac{\delta h(x)}{\delta_K h(x')}\Delta h(x')\,dx' + \int \frac{\delta h(x)}{\delta_K\phi(x')}\Delta\phi(x')\,dx' + \frac{1}{2}\iint \frac{\delta^2 h(x)}{\delta_K h(x')\delta_K h(x'')}\Delta h(x')\Delta h(x'')\,dx'dx''$$

$$+ \iint \frac{\delta^2 h(x)}{\delta_K\phi(x')\delta_K h(x'')}\Delta\phi(x')\Delta h(x'')\,dx'dx'' + \frac{1}{2}\iint \frac{\delta^2 h(x)}{\delta_K\phi(x')\delta_K\phi(x'')}\Delta\phi(x')\Delta\phi(x'')\,dx'dx'' + \ldots \tag{60a}$$

and

$$\Delta_K\phi(x) = \int \frac{\delta\phi(x)}{\delta_K h(x')}\Delta h(x')\,dx' + \int \frac{\delta\phi(x)}{\delta_K\phi(x')}\Delta\phi(x')\,dx' + \frac{1}{2}\iint \frac{\delta^2\phi(x)}{\delta_K h(x')\delta_K h(x'')}\Delta h(x')\Delta h(x'')\,dx'dx''$$

$$+ \iint \frac{\delta^2\phi(x)}{\delta_K\phi(x')\delta_K h(x'')}\Delta\phi(x')\Delta h(x'')\,dx'dx'' + \frac{1}{2}\iint \frac{\delta^2\phi(x)}{\delta_K\phi(x')\delta_K\phi(x'')}\Delta\phi(x')\Delta\phi(x'')\,dx'dx'' + \ldots\,, \tag{60b}$$

where the derivatives are calculated as the unconstrained first and second derivatives of

$$\left(h_K^0[h,\phi], \phi_K^0[h,\phi]\right) = \left( h(x)\frac{Ah_0}{\int h(x')\,dx'},\ \phi(x)\frac{A\phi_0 h_0}{\int \phi(x')h(x')\frac{Ah_0}{\int h(x'')\,dx''}\,dx'} \right). \tag{61}$$

Eq.(61) is the degree-zero K-homogeneous extension of $(h_K, \phi_K)$ satisfying Eqs.(21) and (22) (see [2]). Eq.(60) is nothing else than the Taylor expansion of Eq.(61). Note that terms containing $\dfrac{\delta}{\delta_K \phi}$ vanish in Eq.(60a), since $h_K^0[h,\phi]$ actually does not have a dependence on $\phi(x)$. We mention here that the first-order variations $\delta_K h(x)$ and $\delta_K \phi(x)$ (i.e., $\Delta_{\bar{K}} h(x)$ and $\Delta_{\bar{K}} \phi(x)$), satisfying

$$\int \delta_K \phi(x)\,h(x)\,dx + \int \phi(x)\,\delta_K h(x)\,dx = 0 \tag{62}$$

instead of the full constraint Eq.(58), are given by

$$\delta_K h(x) = \int \frac{\delta h(x)}{\delta_K h(x')}\,\delta h(x')\,dx' + \int \frac{\delta h(x)}{\delta_K \phi(x')}\,\delta\phi(x')\,dx' = \int \left(\delta(x-x') - \frac{h(x)}{N}\right)\delta h(x')\,dx' \tag{63a}$$

and

$$\delta_K \phi(x) = \int \frac{\delta\phi(x)}{\delta_K \phi(x')}\,\delta\phi(x')\,dx' + \int \frac{\delta\phi(x)}{\delta_K h(x')}\,\delta h(x')\,dx'$$

$$= \int \left(\delta(x-x') - \frac{\phi(x)h(x')}{B}\right)\delta\phi(x')\,dx' + \int -\phi(x)\left(\frac{\phi(x')}{B} - \frac{1}{N}\right)\delta h(x')\,dx'. \tag{63b}$$

Inserting Eqs.(60) into Eq.(59), the linear operators acting on $\begin{pmatrix}\Delta h(x') \\ \Delta \phi(x')\end{pmatrix}$, namely,

$$\left( \int dx'\left(\delta(x-x') - \frac{h(x)}{N}\right),\ 0 \right), \tag{64a}$$

and

$$\left( \int dx'\,-\phi(x)\left(\frac{\phi(x')}{B} - \frac{1}{N}\right),\ \int dx'\left(\delta(x-x') - \frac{\phi(x)h(x')}{B}\right) \right), \tag{64b}$$

can be carried over to the kernels in Eq.(59). [We emphasize here that any $(\Delta_K h(x), \Delta_K \phi(x))$ can be taken in the form Eq.(60) due to the construction of $(h_K[h,\phi], \phi_K[h,\phi])$.] By this, we obtain an expression where the variations $\Delta h(x)$ and $\Delta \phi(x)$ are unconstrained:

$$F_T[h+\Delta h, \phi+\Delta\phi] = F_T[h,\phi] + \frac{1}{2}\iint\left\{\left(\frac{\delta^2 F_T[h,\phi]}{\delta_K h(x)\,\delta_K h(x')}\right)^* + \mu_2 2B\left(\frac{\phi(x')}{B}-\frac{1}{N}\right)\left(\frac{\phi(x)}{B}-\frac{1}{N}\right)\right\}\Delta h(x)\Delta h(x')\,dxdx'$$

$$+ \iint\left\{\left(\frac{\delta^2 F_T[h,\phi]}{\delta_K \phi(x)\,\delta_K h(x')}\right)^* - \mu_2\left(\delta(x-x') - \frac{h(x)\phi(x')}{B}\right)\right\}\Delta\phi(x)\Delta h(x')\,dxdx'$$

$$+ \frac{1}{2}\iint\left(\frac{\delta^2 F_T[h,\phi]}{\delta_K \phi(x)\,\delta_K \phi(x')}\right)^* \Delta\phi(x)\Delta\phi(x')\,dxdx' + \text{higher-order terms}, \qquad (65)$$

with

$$\left(\frac{\delta^2 F_T[h,\phi]}{\delta_K h(x')\,\delta_K h(x)}\right)^* = \frac{\delta^2 F_T[h,\phi]}{\delta h(x')\,\delta h(x)} - \frac{1}{N}\int h(x'')\frac{\delta^2 F_T[h,\phi]}{\delta h(x'')\,\delta h(x)}dx''$$

$$-\frac{1}{N}\int h(x'')\frac{\delta^2 F_T[h,\phi]}{\delta h(x')\,\delta h(x'')}dx'' + \frac{1}{N^2}\iint h(x'')h(x''')\frac{\delta^2 F_T[h,\phi]}{\delta h(x'')\,\delta h(x''')}dx''dx'''$$

$$-\left(\frac{\phi(x')}{B}-\frac{1}{N}\right)\int \phi(x'')\frac{\delta^2 F_T[h,\phi]}{\delta\phi(x'')\,\delta h(x)}dx'' + \frac{1}{N}\left(\frac{\phi(x')}{B}-\frac{1}{N}\right)\iint h(x'')\phi(x''')\frac{\delta^2 F_T[h,\phi]}{\delta\phi(x'')\,\delta h(x''')}dx''dx'''$$

$$-\left(\frac{\phi(x)}{B}-\frac{1}{N}\right)\int \phi(x'')\frac{\delta^2 F_T[h,\phi]}{\delta h(x')\,\delta\phi(x'')}dx'' + \frac{1}{N}\left(\frac{\phi(x)}{B}-\frac{1}{N}\right)\iint h(x'')\phi(x''')\frac{\delta^2 F_T[h,\phi]}{\delta h(x'')\,\delta\phi(x''')}dx''dx'''$$

$$+\left(\frac{\phi(x)}{B}-\frac{1}{N}\right)\left(\frac{\phi(x')}{B}-\frac{1}{N}\right)\iint \phi(x'')\phi(x''')\frac{\delta^2 F_T[h,\phi]}{\delta\phi(x'')\,\delta\phi(x''')}dx''dx''', \qquad (66)$$

$$\left(\frac{\delta^2 F_T[h,\phi]}{\delta_K h(x')\,\delta_K \phi(x)}\right)^* = \frac{\delta^2 F_T[h,\phi]}{\delta h(x')\,\delta\phi(x)} - \frac{h(x)}{B}\int \phi(x'')\frac{\delta^2 F_T[h,\phi]}{\delta h(x')\,\delta\phi(x'')}dx''$$

$$-\frac{1}{N}\int h(x'')\frac{\delta^2 F_T[h,\phi]}{\delta h(x'')\,\delta\phi(x)}dx'' + \frac{h(x)}{NB}\iint \phi(x'')h(x''')\frac{\delta^2 F_T[h,\phi]}{\delta h(x''')\,\delta\phi(x'')}dx''dx'''$$

$$-\left(\frac{\phi(x')}{B}-\frac{1}{N}\right)\int \phi(x'')\frac{\delta^2 F_T[h,\phi]}{\delta\phi(x'')\,\delta\phi(x)}dx'' + \frac{h(x)}{B}\left(\frac{\phi(x')}{B}-\frac{1}{N}\right)\iint \phi(x'')\phi(x''')\frac{\delta^2 F_T[h,\phi]}{\delta\phi(x'')\,\delta\phi(x''')}dx''dx''',$$

$$(67)$$

and

$$\left(\frac{\delta^2 F_T[h,\phi]}{\delta_K\phi(x')\delta_K\phi(x)}\right)^* = \frac{\delta^2 F_T[h,\phi]}{\delta\phi(x')\delta\phi(x)} - \frac{h(x)}{B}\int\phi(x'')\frac{\delta^2 F_T[h,\phi]}{\delta\phi(x')\delta\phi(x'')}dx''$$

$$-\frac{h(x')}{B}\int\phi(x'')\frac{\delta^2 F_T[h,\phi]}{\delta\phi(x'')\delta\phi(x)}dx'' + \frac{h(x)h(x')}{B^2}\iint\phi(x'')\phi(x''')\frac{\delta^2 F_T[h,\phi]}{\delta\phi(x'')\delta\phi(x''')}dx''dx''' \ . \tag{68}$$

The emerging higher-order terms in Eq.(65) due to the full expansion Eq.(60) are irrelevant with respect to equilibrium analysis, just as the original higher-order terms, as pointed out previously.

As can be observed, the general expression Eq.(65) reduces significantly in the case of the special equilibrium considered in Sec.III, since, e.g.,

$$\left(\frac{\phi_0}{B} - \frac{1}{N}\right) = 0 \ . \tag{69}$$

It will give back Eq.(38). Note that the term containing $\frac{h(x)\phi(x')}{B}$, e.g., is what ensures the cancellation of $\Delta h_0$ and $\Delta\phi_0$ of the Furier expansions of $\Delta h(x)$ and $\Delta\phi(x)$. It has to be underlined that in the general case, $\mu_2$ appears not only beside $\Delta\phi(x)\Delta h(x')$ but also beside $\Delta h(x)\Delta h(x')$.

### B. Accounting for all effects due to constraints through constrained derivatives

As we have seen, freeing the variations from the constraints naturally leads to the appearance of constrained derivatives – as defined according to Eqs.(11) and (53). However, there is more in their concept regarding the analysis of functionals under constraints.

Observe that constrained second derivatives defined according to Eq.(50) can be written as

$$\frac{\delta^2 A[\rho]}{\delta_C\rho(x')\delta_C\rho(x)} = \frac{\delta^2 A[\rho_C[\rho]]}{\delta\rho(x')\delta\rho(x)}\bigg|_{\rho=\rho_C} = \frac{\delta}{\delta\rho(x')}\int \frac{\delta A}{\delta\rho(x'')}[\rho_C[\rho]]\frac{\delta\rho_C[\rho](x'')}{\delta\rho(x)}dx''\bigg|_{\rho=\rho_C}$$

$$=\iint \frac{\delta^2 A}{\delta\rho(x''')\delta\rho(x'')}[\rho_C[\rho]]\frac{\delta\rho_C[\rho](x''')}{\delta\rho(x')}\frac{\delta\rho_C[\rho](x'')}{\delta\rho(x)}dx''dx'''\bigg|_{\rho=\rho_C} + \int \frac{\delta A}{\delta\rho(x'')}[\rho_C[\rho]]\frac{\delta^2\rho_C[\rho](x'')}{\delta\rho(x')\delta\rho(x)}dx''\bigg|_{\rho=\rho_C}$$

$$=\iint \frac{\delta^2 A[\rho]}{\delta\rho(x''')\delta\rho(x'')}\frac{\delta\rho(x''')}{\delta_C\rho(x')}\frac{\delta\rho(x'')}{\delta_C\rho(x)}dx''dx''' + \int \frac{\delta A[\rho]}{\delta\rho(x'')}\frac{\delta^2\rho(x'')}{\delta_C\rho(x')\delta_C\rho(x)}dx''$$

$$=\left(\frac{\delta^2 A[\rho]}{\delta_C\rho(x')\delta_C\rho(x)}\right)^* + \int \frac{\delta A[\rho]}{\delta\rho(x'')}\frac{\delta^2\rho(x'')}{\delta_C\rho(x')\delta_C\rho(x)}dx'' \ . \tag{70}$$

Applying a further differentiation with respect to $\rho(x)$ in the second line of Eq.(70) gives the third-order constrained derivative,

$$\frac{\delta^3 A[\rho]}{\delta_C\rho(x'')\delta_C\rho(x')\delta_C\rho(x)} = \left(\frac{\delta^3 A[\rho]}{\delta_C\rho(x'')\delta_C\rho(x')\delta_C\rho(x)}\right)^* + \int \frac{\delta^2 A[\rho]}{\delta\rho(x''')\delta\rho(x''')}\frac{\delta^2\rho(x''')}{\delta_C\rho(x'')\delta_C\rho(x')}\frac{\delta\rho(x''')}{\delta_C\rho(x)}dx'''$$

$$+\int \frac{\delta^2 A[\rho]}{\delta\rho(x''')\delta\rho(x''')}\frac{\delta^2\rho(x''')}{\delta_C\rho(x')\delta_C\rho(x)}\frac{\delta\rho(x''')}{\delta_C\rho(x')}dx''' + \int \frac{\delta^2 A[\rho]}{\delta\rho(x''')\delta\rho(x''')}\frac{\delta^2\rho(x''')}{\delta_C\rho(x')\delta_C\rho(x)}\frac{\delta\rho(x''')}{\delta_C\rho(x')}dx'''$$

$$+\int \frac{\delta A[\rho]}{\delta\rho(x''')}\frac{\delta^3\rho(x''')}{\delta_C\rho(x'')\delta_C\rho(x')\delta_C\rho(x)}dx''' \ , \tag{71}$$

and so on. In Eq.(71), the first term on the right side is defined by

$$\left(\frac{\delta^n A[\rho]}{\delta_C\rho(x)\cdots\delta_C\rho(x^{(n)})}\right)^* = \int\cdots\int \frac{\delta^n A[\rho]}{\delta\rho(x_1)\cdots\delta\rho(x_n)}\frac{\delta\rho(x_1)}{\delta_C\rho(x)}\cdots\frac{\delta\rho(x_n)}{\delta_C\rho(x^{(n)})}dx_1\cdots dx_n \ . \tag{72}$$

Note that for $\frac{\delta A[\rho]}{\delta_C\rho(x)}$, Eq.(56) holds; that is, $\frac{\delta A[\rho]}{\delta_C\rho(x)} = \left(\frac{\delta A[\rho]}{\delta_C\rho(x)}\right)^*$.

Now, insert Eq.(51) into $A[\rho]$'s Taylor expansion (44), and collect the terms of same order in $\Delta\rho(x)$, to find, with the help of Eqs.(70) and (71),

$$A[\rho+\Delta_C\rho] = A[\rho] + \int \frac{\delta A[\rho]}{\delta_C\rho(x)}\Delta\rho(x)dx + \frac{1}{2!}\iint \frac{\delta^2 A[\rho]}{\delta_C\rho(x)\delta_C\rho(x')}\Delta\rho(x)\Delta\rho(x')dxdx'$$

$$+ \frac{1}{3!}\iiint \frac{\delta^3 A[\rho]}{\delta_C\rho(x)\delta_C\rho(x')\delta_C\rho(x'')}\Delta\rho(x)\Delta\rho(x')\Delta\rho(x'')dxdx'dx'' + \ldots \tag{73}$$

By calculating the expressions corresponding to Eqs.(70) and (71) for higher-order constrained derivatives, it is easy to see that Eq.(73) can be continued for higher-order terms.

At a stationary (or critical) point $\rho(x)$, $\frac{\delta A[\rho]}{\delta_C \rho(x)} = 0$ [14]; thus, the first term in Eq.(73) vanishes. For the second-order term of Eq.(73), we can establish a result analogous to Eq.(40) with the help of Eq.(54).

*Theorem – necessary condition for a local extremum.* If at a critical point $\rho(x)$ of a functional $A$, there is a local minimum (maximum) of $A$,

$$\iint \frac{\delta^2 A[\rho]}{\delta_C \rho(x) \delta_C \rho(x')} \Delta\rho(x) \Delta\rho(x') dx dx' \geq (\leq) \, 0 \qquad \text{for all } \Delta\rho(x)\text{'s}. \qquad (74a)$$

With strict inequalities, we obtain a sufficient condition for the existence of local extremum:

*Theorem – sufficient condition for a local extremum.* If at a critical point $\rho(x)$ of a functional $A$,

$$\iint \frac{\delta^2 A[\rho]}{\delta_C \rho(x) \delta_C \rho(x')} \Delta\rho(x) \Delta\rho(x') dx dx' \geq p > 0 \, (\leq p < 0) \qquad \text{for all* nonzero } \Delta\rho(x)\text{'s}, \qquad (74b)$$

then there is a local minimum (maximum) of $A$. $p$ is an arbitrarily small (in absolute value) positive (negative) constant, independent of $\Delta\rho(x)$. In other words, the infimum (supremum) of the constrained second differential has to be greater (less) than zero to have a local minimum (maximum). This is similarly so in the unconstrained case, and also in the case of Eq.(41) [13]. The presence of $p$ in Eq.(74b) is important; it is to rule out cases of sequences of positive (negative) second differentials tending to zero with $\Delta\rho(x)$. Because of this, the domain of all (nonzero) $\Delta\rho(x)$'s has to be restricted in Eq.(74b) (denoted by all*) to avoid the second differential tending to zero with $\Delta\rho(x) \to 0$ – which could give a zero infimum (supremum). This can be achieved formally, e.g., by restricting $\Delta\rho(x)$ to be of $\|\Delta\rho(x)\| = 1$. But note that this is not a drastic restriction, since any $\Delta\rho(x)$ can be obtained by multiplying a $(\Delta\rho(x))_1$ of norm one by some positive constant; that is, we do not have to account for a constraint $\|\Delta\rho(x)\|=1$ on the variations $\Delta\rho(x)$.

*Proof.* Inserting Eq.(47) into Eq.(70) gives

$$\frac{\delta^2 A[\rho]}{\delta_C\rho(x')\delta_C\rho(x)} = \left(\frac{\delta^2 A[\rho]}{\delta_C\rho(x')\delta_C\rho(x)}\right)^* + \mu\int \frac{\delta C[\rho]}{\delta\rho(x'')} \frac{\delta^2\rho(x'')}{\delta_C\rho(x')\delta_C\rho(x)} dx'' \quad . \tag{75}$$

Compare Eq.(74) (with Eq.(75) inserted) with Eq.(54) to see that all we have to prove is

$$\int \frac{\delta C[\rho]}{\delta\rho(x'')} \frac{\delta^2\rho(x'')}{\delta_C\rho(x')\delta_C\rho(x)} dx'' = -\int \frac{\delta^2 C[\rho]}{\delta\rho(x''')\delta\rho(x'')} \frac{\delta\rho(x''')}{\delta_C\rho(x')} \frac{\delta\rho(x'')}{\delta_C\rho(x)} dx'' \quad . \tag{76}$$

Eq.(76) can be proved by inserting the expansion Eq.(51) into Eq.(46), and noticing that since the variation $\Delta\rho(x)$ is now unconstrained, terms of the same order in $\Delta\rho(x)$ have to cancel each other. Eq.(76) emerges from the second-order terms cancelling each other. (Alternatively, $\frac{\delta C[\rho_C[\rho]]}{\delta\rho(x)}=0$ may also be differentiated to obtain Eq.(76).) □

It is worth mentioning here that expressions of higher-order similar to Eq.(76) can be obtained in the way Eq.(76) has been obtained.

In practice, Eq.(74b) can be applied in the way Eq.(40) is usually applied – by examining the eigenvalue spectrum of the second derivative [16,17]. In the presence of constraints, the eigenvalue equation becomes

$$\int \frac{\delta^2 A[\rho]}{\delta_C\rho(x)\delta_C\rho(x')} \Delta\rho(x') dx' = \lambda \Delta\rho(x) \quad . \tag{77}$$

If all the eigenvalues are greater (less) than some positive (negative) number $p$ arbitrarily close to zero, there is a local minimum (maximum) at the examined stationary point. (As explained below Eq.(74b), $p$ is needed to exclude zero being a cluster point of the $\lambda$'s.) This can be proved on the basis of Eq.(74b) completely analogously to the unconstrained case. A local minimum usually represents a stable equilibrium, while the other cases imply an instable, or metastable, equilibrium. This method has been applied by Uline and Corti in [12], in the stability analysis of droplet growth in supercooled vapors.

Recently, there has been much interest in the problem of accounting for constraints in the analysis of stationary points, both in the infinite-dimensional [18-20] and in the finite-dimensional case [21,22]. It has been established that in the presence of constraints, the eigenvalues of

$$\int \left( \frac{\delta^2 A[\rho]}{\delta\rho(x)\delta\rho(x')} - \mu \frac{\delta^2 C[\rho]}{\delta\rho(x)\delta\rho(x')} \right) \Delta\rho(x') dx' = \lambda \Delta\rho(x)$$

give us a tool to determine if at a critical point, there is a local minimum or maximum, or neither of them. However, although it is true that if $\lambda \geq p > 0$ for all $\lambda$ at a given critical point, then there is a local minimum, there can still be a local minimum if there is only one negative $\lambda$ [19]. For this case, Vogel [19] has proved a criterion to determine whether there indeed is a local minimum, or not. Examining the eigenvalue spectrum of Eq.(77) presents an alternative way to decide about the nature of a critical point.

Turning to the concrete example of a two-variable functional with constraints (21) and (22), the corresponding constrained second derivatives emerge as

$$\frac{\delta^2 A[h,\phi]}{\delta_K h(x') \delta_K h(x)} = \left( \frac{\delta^2 A[h,\phi]}{\delta_K h(x') \delta_K h(x)} \right)^* + \int \frac{\delta A[h,\phi]}{\delta h(x'')} \frac{\delta^2 h(x'')}{\delta_K h(x') \delta_K h(x)} dx'' + \int \frac{\delta A[h,\phi]}{\delta \phi(x'')} \frac{\delta^2 \phi(x'')}{\delta_K h(x') \delta_K h(x)} dx''$$

$$= \left( \frac{\delta^2 A[h,\phi]}{\delta_K h(x') \delta_K h(x)} \right)^* - \frac{1}{N}\left( \frac{\delta A[h,\phi]}{\delta_K h(x)} + \frac{\delta A[h,\phi]}{\delta_K h(x')} \right) + 2\left( \frac{\phi(x')}{B} - \frac{1}{N} \right)\left( \frac{\phi(x)}{B} - \frac{1}{N} \right) \int \phi(x'') \frac{\delta A[h,\phi]}{\delta \phi(x'')} dx'' ,$$

(78)

$$\frac{\delta^2 A[h,\phi]}{\delta_K h(x') \delta_K \phi(x)} = \left( \frac{\delta^2 A[h,\phi]}{\delta_K h(x') \delta_K \phi(x)} \right)^* + \int \frac{\delta A[h,\phi]}{\delta h(x'')} \frac{\delta^2 h(x'')}{\delta_K h(x') \delta_K \phi(x)} dx'' + \int \frac{\delta A[h,\phi]}{\delta \phi(x'')} \frac{\delta^2 \phi(x'')}{\delta_K h(x') \delta_K \phi(x)} dx''$$

$$= \left( \frac{\delta^2 A[h,\phi]}{\delta_K h(x') \delta_K \phi(x)} \right)^* - \left( \frac{\phi(x')}{B} - \frac{1}{N} \right) \frac{\delta A[h,\phi]}{\delta_K \phi(x)} - \left( \delta(x-x') - \frac{h(x)\phi(x')}{B} \right) \frac{1}{B} \int \phi(x'') \frac{\delta A[h,\phi]}{\delta \phi(x'')} dx'' ,$$

(79)

and

$$\frac{\delta^2 A[h,\phi]}{\delta_K \phi(x') \delta_K \phi(x)} = \left(\frac{\delta^2 A[h,\phi]}{\delta_K \phi(x') \delta_K \phi(x)}\right)^* + \int \frac{\delta A[h,\phi]}{\delta h(x'')} \frac{\delta^2 h(x'')}{\delta_K \phi(x') \delta_K \phi(x)} dx'' + \int \frac{\delta A[h,\phi]}{\delta \phi(x'')} \frac{\delta^2 \phi(x'')}{\delta_K \phi(x') \delta_K \phi(x)} dx''$$

$$= \left(\frac{\delta^2 A[h,\phi]}{\delta_K \phi(x') \delta_K \phi(x)}\right)^* - \frac{h(x')}{B} \frac{\delta A[h,\phi]}{\delta_K \phi(x)} - \frac{h(x)}{B} \frac{\delta A[h,\phi]}{\delta_K \phi(x')} \quad . \tag{80}$$

The $\dfrac{\delta^2 \rho(x'')}{\delta_K \rho(x') \delta_K \rho(x)}$ terms in Eqs.(78), (79), and (80) are worth giving explicitly:

$$\frac{\delta^2 h(x'')}{\delta_K h(x') \delta_K h(x)} = -\frac{1}{N}\left(\frac{\delta h(x'')}{\delta_K h(x)} + \frac{\delta h(x'')}{\delta_K h(x')}\right)$$

$$= -\frac{1}{N}\left(\frac{\delta h(x'')}{\delta_N h(x)} + \frac{\delta h(x'')}{\delta_N h(x')}\right) = -\frac{1}{N}\left(\delta(x''-x) + \delta(x''-x') - 2\frac{h(x'')}{N}\right), \tag{81a}$$

$$\frac{\delta^2 \phi(x'')}{\delta_K h(x') \delta_K h(x)} = -\frac{1}{N}\left(\frac{\delta \phi(x'')}{\delta_K h(x)} + \frac{\delta \phi(x'')}{\delta_K h(x')}\right) + 2\left(\frac{\phi(x')}{B} - \frac{1}{N}\right)\left(\frac{\phi(x)}{B} - \frac{1}{N}\right)\phi(x'')$$

$$= \frac{\phi(x)}{B}\left(\frac{\phi(x')}{B} - \frac{1}{N}\right)\phi(x'') + \frac{\phi(x')}{B}\left(\frac{\phi(x)}{B} - \frac{1}{N}\right)\phi(x'') \quad , \tag{81b}$$

$$\frac{\delta^2 h(x'')}{\delta_K h(x') \delta_K \phi(x)} = -\left(\frac{\phi(x')}{B} - \frac{1}{N}\right)\frac{\delta h(x'')}{\delta_K \phi(x)} = 0 \quad , \tag{82a}$$

$$\frac{\delta^2 \phi(x'')}{\delta_K h(x') \delta_K \phi(x)} = -\left(\frac{\phi(x')}{B} - \frac{1}{N}\right)\frac{\delta \phi(x'')}{\delta_K \phi(x)} - \left(\delta(x-x') - \frac{h(x)\phi(x')}{B}\right)\frac{\phi(x'')}{B} \quad , \tag{82b}$$

and

$$\frac{\delta^2 h(x'')}{\delta_K \phi(x') \delta_K \phi(x)} = -\frac{h(x')}{B}\frac{\delta h(x'')}{\delta_K \phi(x)} - \frac{h(x)}{B}\frac{\delta h(x'')}{\delta_K \phi(x')} = 0 \quad , \tag{83a}$$

$$\frac{\delta^2 \phi(x'')}{\delta_K \phi(x') \delta_K \phi(x)} = -\frac{h(x')}{B}\frac{\delta \phi(x'')}{\delta_K \phi(x)} - \frac{h(x)}{B}\frac{\delta \phi(x'')}{\delta_K \phi(x')} \quad . \tag{83b}$$

The (unconstrained) second derivatives, of course, vanish for these special cases (which are obtained also if one substitutes $A[h,\phi] = h(x)$, and $A[h,\phi] = \phi(x)$, into Eqs.(78)-(80)).

The condition of (total) instability of the equilibrium state in the case of the free-energy functional $F_T[h,\phi]$ will be

$$\frac{1}{2}\iint \frac{\delta^2 F_T[h,\phi]}{\delta_K h(x')\delta_K h(x)}\Delta h(x)\Delta h(x')\,dxdx' + \iint \frac{\delta^2 F_T[h,\phi]}{\delta_K h(x')\delta_K \phi(x)}\Delta\phi(x)\Delta h(x')\,dxdx'$$

$$+\frac{1}{2}\iint \frac{\delta^2 F_T[h,\phi]}{\delta_K \phi(x')\delta_K \phi(x)}\Delta\phi(x)\Delta\phi(x')\,dxdx' < 0$$

[for all $(\Delta h(x), \Delta\phi(x'))$ s]. (84)

When applying this result for the special case of equilibrium considered in [4], and in the preceding section, many of the terms of Eqs.(78)–(80) vanish, giving

$$\iint \frac{\delta^2 F_T[h,\phi]}{\delta_K h(x')\delta_K h(x)}\Delta h(x)\Delta h(x')\,dxdx' = \frac{\partial^2 f_T(h_0,\phi_0)}{\partial h^2}\left(\int (\Delta h(x))^2\,dx - \frac{1}{A}\left(\int \Delta h(x)\,dx\right)^2\right), \quad (85)$$

$$\iint \frac{\delta^2 F_T[h,\phi]}{\delta_K h(x')\delta_K \phi(x)}\Delta\phi(x)\Delta h(x')\,dxdx' = \frac{\partial^2 f_T(h_0,\phi_0)}{\partial h\,\partial\phi}\left(\int \Delta\phi(x)\Delta h(x)\,dx - \frac{1}{A}\int \Delta\phi(x)\,dx\int \Delta h(x)\,dx\right)$$

$$-\mu_2\left(\int \Delta\phi(x)\Delta h(x)\,dx - \frac{1}{A}\int \Delta\phi(x)\,dx\int \Delta h(x)\,dx\right), \quad (86)$$

and

$$\iint \frac{\delta^2 F_T[h,\phi]}{\delta_K \phi(x')\delta_K \phi(x)}\Delta\phi(x)\Delta\phi(x')\,dxdx' = \frac{\partial^2 f_T(h_0,\phi_0)}{\partial \phi^2}\left(\int (\Delta\phi(x))^2\,dx - \frac{1}{A}\left(\int \Delta\phi(x)\,dx\right)^2\right), \quad (87)$$

with $\mu_2 = \frac{1}{h_0}\frac{\partial f_T(h_0,\phi_0)}{\partial \phi}$. Then inserting the Furier expansions Eqs.(31) into the above differentials, and applying Eq.(84) give back Eq.(39). It can be seen that relying on constrained second derivatives, directly leads to the final result, without any further considerations regarding the proper account for constraints. There is no need to consider additional terms coming from the first differential due to the constraints, and to account for constrained variations.

It can be concluded that the constrained second derivatives defined according to Eq.(50) provide the proper ground for the analysis of functionals (including the analysis of stationary points) under constraints. (This answers the question raised in Section II, too.) On the basis of Eq.(73), it may be not too bold to expect this finding to be valid also for higher-order constrained derivatives, defined by Eq.(50). As seen, eventually this result has been obtained on the basis of the idea behind Eq.(15). The reason that idea gave a definition in Sec.II that is different from Eq.(19) in the case of the *N*-conservation constraint is the special nature of linear constraints – a kind of degeneracy manifested also with respect to the issue of simultaneous constraints [23], and with respect to the origination of *K*-constrained derivatives as Gâteaux derivatives along *K*-conserving paths [14]. For, $\Delta_N \rho(x)$ (given by Eq.(11b)) itself fulfils the constraint Eq.(9), leading to a different *N*-constrained second derivative than Eq.(19). However, taking the full expansion Eq.(51) for $\Delta_N \rho(x)$ into account, formally also yields the same definition as Eq.(19). Having mentioned the Gâteaux kind of definition of derivatives along *K*-conserving paths [14], it is worth giving the corresponding formula for the (nonlinear) constraint of Eqs.(21)-(22),

$$D_G\big|_K (A)\,[(h,\phi);(\Delta_K h, \Delta_K \phi)] = \lim_{\varepsilon \to 0} \frac{1}{\varepsilon}\left\{ A\left[ h + \varepsilon \Delta_K h, \frac{B}{\int (\phi + \varepsilon \Delta_K \phi)(h + \varepsilon \Delta_K h)} (\phi + \varepsilon \Delta_K \phi) \right] - A[h,\phi] \right\}. \quad (88)$$

It is interesting to examine the second derivatives obtained by two successive *K*-constrained differentiations – i.e., the case of Eq.(14). The second *K*-constrained derivatives (for the constraint of Eqs.(21)-(22)) will be

$$\frac{\delta}{\delta_K h(x')} \frac{\delta A[h,\phi]}{\delta_K h(x)} = \left(\frac{\delta^2 A[h,\phi]}{\delta_K h(x') \delta_K h(x)}\right)^* - \frac{1}{N}\frac{\delta A[h,\phi]}{\delta_N h(x')} + 2\frac{\phi(x)}{B}\left(\frac{\phi(x')}{B} - \frac{1}{N}\right)\int \phi(x'') \frac{\delta A[h,\phi]}{\delta \phi(x'')} dx''$$

$$= \frac{\delta^2 A[h,\phi]}{\delta_K h(x') \delta_K h(x)} + \frac{1}{N}\frac{\delta A[h,\phi]}{\delta_K h(x)} + \frac{\phi(x)}{B}\left(\frac{\phi(x')}{B} - \frac{1}{N}\right)\int \phi(x'') \frac{\delta A[h,\phi]}{\delta \phi(x'')} dx'' \,, \quad (89)$$

$$\frac{\delta}{\delta_K \phi(x')} \frac{\delta A[h,\phi]}{\delta_K h(x)} = \left(\frac{\delta^2 A[h,\phi]}{\delta_K \phi(x') \delta_K h(x)}\right)^* - \left(\frac{\phi(x)}{B} - \frac{1}{N}\right)\frac{\delta A[h,\phi]}{\delta_K \phi(x')} - \left(\delta(x-x') - \frac{h(x')\phi(x)}{B}\right)\frac{1}{B}\int \phi(x'') \frac{\delta A[h,\phi]}{\delta \phi(x'')} dx''$$

$$= \frac{\delta^2 A[h,\phi]}{\delta_K h(x) \delta_K \phi(x')} \,, \quad (90)$$

$$\frac{\delta}{\delta_K h(x')} \frac{\delta A[h,\phi]}{\delta_K \phi(x)} = \left(\frac{\delta^2 A[h,\phi]}{\delta_K h(x') \delta_K \phi(x)}\right)^* - \left(\delta(x-x') - \frac{h(x)\phi(x')}{B}\right)\frac{1}{B}\int \phi(x'') \frac{\delta A[h,\phi]}{\delta \phi(x'')} dx''$$

$$= \frac{\delta}{\delta_K \phi(x)} \frac{\delta A[h,\phi]}{\delta_K h(x')} + \left(\frac{\phi(x')}{B} - \frac{1}{N}\right) \frac{\delta A[h,\phi]}{\delta_K \phi(x)} , \qquad (91)$$

and

$$\frac{\delta}{\delta_K \phi(x')} \frac{\delta A[h,\phi]}{\delta_K \phi(x)} = \left(\frac{\delta^2 A[h,\phi]}{\delta_K \phi(x') \delta_K \phi(x)}\right)^* - \frac{h(x)}{B} \frac{\delta A[h,\phi]}{\delta_K \phi(x')} . \qquad (92)$$

Eqs.(89)-(92) show that not only do the derivatives (89) and (92) not retain the symmetry in $(x',x)$ of a symmetric $\frac{\delta^2 A[h,\phi]}{\delta h(x') \delta h(x)}$, and $\frac{\delta^2 A[h,\phi]}{\delta \phi(x') \delta \phi(x)}$, respectively, but even the order of differentiation with respect to $h(x)$ and $\phi(x)$ becomes relevant.

Finally, we mention that the arguments of this study can be applied to finite-dimensional vector spaces, where constrained derivatives can be introduced, too [23]. In that case, Eq.(74a), e.g., takes the form

$$\sum_{i,j=1}^{n} \frac{\partial^2 a(x_1,...,x_n)}{\partial_C x_i \partial_C x_j} \Delta x_i \Delta x_j \geq (\leq) 0 . \qquad (93)$$

## V. Summary

It has been shown that constrained second derivatives, defined according to Eq.(50), incorporate all second-order effects due to constraints; consequently, constrained second derivatives provide the proper tool for physics for the stability analysis of equilibria under conservation constraints. In the presence of constraints, Eq.(74) gives the proper generalization of the well-known condition [Eq.(40)] for the existence of a local extremum. More generally, it can be concluded on the basis of Eq.(73) that under constraints, the unconstrained derivatives of order *m* have to be replaced by the corresponding constrained derivatives (50) in problems based on the Taylor expansion of the functional in question. For the physically important type of constraints of Eqs.(21)-(22) (for which the constrained second derivatives are given in Eqs.(78)-(80)), it has been demonstrated how the presented theory works, showing also how the stability condition obtained by Clarke for a special case of equilibrium in his thin-film dynamical model [4,5] emerges.

## Appendix: On the choice of the mapping $\rho_C[\rho]$

The most general form for a constrained first derivative, in the case of the norm-conserving constraint (9), is [14]

$$\frac{\delta A[\rho]}{\delta_N \rho(x)} = \frac{\delta A[\rho]}{\delta \rho(x)} - \int u(x') \frac{\delta A[\rho]}{\delta \rho(x')} dx' , \qquad (A1)$$

where $u(x)$ is an arbitrary function that integrates to one. Eq.(A1) can be obtained from [23]

$$\rho_N[\rho] = \rho(x) - u(x)\left(\int \rho(x')dx' - N\right) , \qquad (A2)$$

inserted into Eq.(50). The constrained derivatives emerging with the use of Eq.(A2) fulfil the K-equality condition, and can be used in Eqs.(73) and (74). As pointed out in [14], the choice $u(x) = \rho(x)\big/\int \rho(x')dx'$ of [3] yields the intuitively appealing property of $\frac{\delta A[\rho]}{\delta_N \rho(x)} = \frac{\delta A[\rho]}{\delta \rho(x)}$ for $N$-independent functionals. In a dynamical theory that does not directly emerge from a variational principle (like the model developed in [5]), the choice of $u(x)$ may have relevance; however, in establishing Eq.(73), and in particular, Eq.(74), $u(x)$ can be chosen arbitrarily, and in practice, that choice can be based on pragmatic considerations. In this Appendix, we will (i) give the constrained second derivatives emerging from Eq.(A2), (ii) examine the special choice of $u(x) = \delta(x - x_0)$, and (iii) show that Eq.(A2) indeed leads to the most general form of higher-order constrained derivatives that can be applied in Eq.(73).

Differentiating $A[\rho_N[\rho]]$ yields

$$\frac{\delta A[\rho_N[\rho]]}{\delta \rho(x)} = \frac{\delta A}{\delta \rho(x)}[\rho_N[\rho]] - \int \left\{ u(x'') + \frac{\delta u(x'')}{\delta \rho(x)}\left(\int \rho(\tilde{x})d\tilde{x} - N\right)\right\} \frac{\delta A}{\delta \rho(x'')}[\rho_N[\rho]] dx'' . \quad (A3)$$

Eq.(A3) gives Eq.(A1) by insertion of $\rho(x)$ with $\int \rho(x)dx = N$. Differentiation of Eq.(A3) gives

$$\frac{\delta^2 A[\rho_N[\rho]]}{\delta\rho(x)\delta\rho(x')} = \frac{\delta^2 A}{\delta\rho(x)\delta\rho(x')}[\rho_N[\rho]] - \int \left\{ u(x'') + \frac{\delta u(x'')}{\delta\rho(x')}\left(\int\rho - N\right)\right\} \frac{\delta^2 A}{\delta\rho(x)\delta\rho(x'')}[\rho_N[\rho]]dx''$$

$$-\int\left\{ u(x'') + \frac{\delta u(x'')}{\delta\rho(x)}\left(\int\rho - N\right)\right\}\left(\frac{\delta^2 A}{\delta\rho(x'')\delta\rho(x')}[\rho_N[\rho]] - \int\left\{u(x''') + \frac{\delta u(x''')}{\delta\rho(x')}\left(\int\rho - N\right)\right\}\frac{\delta^2 A}{\delta\rho(x'')\delta\rho(x''')}[\rho_N[\rho]]dx'''\right)dx''$$

$$-\int\left\{\frac{\delta u(x'')}{\delta\rho(x')} + \frac{\delta u(x'')}{\delta\rho(x)} + \frac{\delta^2 u(x'')}{\delta\rho(x)\delta\rho(x')}\left(\int\rho - N\right)\right\}\frac{\delta A}{\delta\rho(x'')}[\rho_N[\rho]]dx'' \quad . \tag{A4}$$

This then yields the constrained second derivative

$$\frac{\delta^2 A[\rho]}{\delta_N\rho(x)\delta_N\rho(x')} = \frac{\delta^2 A[\rho]}{\delta\rho(x)\delta\rho(x')} - \int u(x'')\left(\frac{\delta^2 A[\rho]}{\delta\rho(x)\delta\rho(x'')} + \frac{\delta^2 A[\rho]}{\delta\rho(x')\delta\rho(x'')}\right)dx'' + \iint u(x'')u(x''')\frac{\delta^2 A[\rho]}{\delta\rho(x'')\delta\rho(x''')}dx''dx'''$$

$$-\int\left(\frac{\delta u(x'')}{\delta\rho(x)} + \frac{\delta u(x'')}{\delta\rho(x')}\right)\frac{\delta A[\rho]}{\delta\rho(x'')}dx'' \quad . \tag{A5}$$

Formally, the simplest choice for $u(x)$ is the Dirac delta function $\delta(x - x_0)$. With the use of it,

$$\frac{\delta A[\rho]}{\delta_N\rho(x)} = \frac{\delta A[\rho]}{\delta\rho(x)} - \frac{\delta A[\rho]}{\delta\rho(x_0)} \tag{A6}$$

emerges as a constrained first derivative [14]. It indeed fulfils the K-equality condition, because any (x-independent) constant added to the derivative $\frac{\delta A[\rho]}{\delta\rho(x)}$ cancels in Eq.(A6). It is interesting to recognize that in density functional theory (DFT) [24,25], this form of constrained derivatives appears explicitly in the basic Euler-Lagrange equation of the theory. In DFT, an energy density functional $E_v[n] = F[n] + \int n(\vec{r})v(\vec{r})d\vec{r}$ is defined whose minimum under the constraint $\int n(\vec{r})d\vec{r} = N$ delivers the ground-state energy of the N-electron system in the scalar external potential $v(\vec{r})$. This minimum principle leads to the Euler-Lagrange equation

$$\frac{\delta F[n]}{\delta n(\vec{r})} + v(\vec{r}) = \mu \tag{A7}$$

for the determination of the ground-state $n(\vec{r})$. Eq.(A7) also gives the external potential as a functional of the ground-state density, $v[n]$, in accordance with the first Hohenberg-Kohn theorem [24], which establishes a one-to-one correspondence between $v(\vec{r})$ (with the arbitrary additive constant fixed by some choice) and $n(\vec{r})$. For electronic potentials $v(\vec{r})$, $v(\infty) = 0$. This gives $\mu = \frac{\delta F[n]}{\delta n(\infty)}$; that is,

$$v(\vec{r})[n] = -\frac{\delta F[n]}{\delta n(\vec{r})} + \frac{\delta F[n]}{\delta n(\infty)} . \tag{A8}$$

Eq.(A8) then gives $v[n]$ uniquely. It also shows that $v[n]$ is a constrained derivative of the density functional $F[n]$,

$$v(\vec{r})[n] = -\frac{\delta F[n]}{\delta_N n(\vec{r})} , \tag{A9}$$

according to Eq.(A6), with $\vec{r}_0 = \infty$.

From Eq.(A5), the constrained second derivative with $u(x) = \delta(x - x_0)$ can be easily obtained:

$$\frac{\delta^2 A[\rho]}{\delta_N \rho(x) \delta_N \rho(x')} = \frac{\delta^2 A[\rho]}{\delta \rho(x) \delta \rho(x')} - \frac{\delta^2 A[\rho]}{\delta \rho(x) \delta \rho(x_0)} - \frac{\delta^2 A[\rho]}{\delta \rho(x_0) \delta \rho(x')} + \frac{\delta^2 A[\rho]}{\delta \rho(x_0) \delta \rho(x_0)} . \tag{A10}$$

This expression seems to be very simple and appealing; however, there are two problems with the use of it. First, in the case of functionals $A[\rho] = \int f(x, \rho(x), \rho^{(1)}(x), \rho^{(2)}(x), ...) dx$, which form frequently appears in physical applications, the delta function enters the second derivative; this means that the last term in Eq.(A10) will contain $\delta(x_0 - x_0)$. Second, in practical calculations, due to their, usually, approximate nature, relying on the value of $\rho(x)$ at one given point may strongly effect the accuracy of the result. For DFT, where the $v \leftrightarrow n$ mapping is defined with $\int n(\vec{r})d\vec{r}$ fixed, we mention that the inverse density response function can be given as

$$\frac{\delta v(\vec{r})}{\delta_N n(\vec{r}')} = -\frac{\delta^2 F[n]}{\delta_N n(\vec{r})\delta_N n(\vec{r}')} \ , \tag{A11}$$

on the basis of Eq.(A9). (Since the exact $F[n]$ is much more complicated than simply having a form $\int f(\vec{r}, n(\vec{r}), ...)d\vec{r}$, the above-mentioned problem does not occur in the exact theory.) Note that if we have an extension of $v[n]$ from the $n_N(\vec{r})$ domain, the $N$-conserving constraint on the differentiation in Eq.(A11) can be relaxed.

We now examine the question as to whether Eq.(A2) indeed embraces all possible $\rho_N[\rho]$'s that give good constrained derivatives of all order, which fulfill the K-equality condition. Notice that there is a wide range of possible $\rho_N[\rho]$'s – i.e., which give a $\rho_N(x)$ for any $\rho(x)$, and become an identity for $\rho_N(x)$'s. These can even involve physics, yielding complicated (possibly non-analytical) forms. The key for answering this question is to observe that $\rho_{\int \rho}[\rho] = \rho$. Differentiation of this relation with respect to $\rho(x)$ gives

$$\frac{\delta \rho_N(x')[\rho]}{\delta \rho(x)} + \frac{\partial \rho_N(x')[\rho]}{\partial N} = \delta(x' - x) \ . \tag{A12}$$

Now, Eq.(A12) can be used in $\frac{\delta A[\rho_N[\rho]]}{\delta \rho(x)} = \int \frac{\delta A}{\delta \rho(x')} \frac{\delta \rho_N(x')[\rho]}{\delta \rho(x)} dx'$ to obtain

$$\frac{\delta A[\rho]}{\delta_N \rho(x)} = \frac{\delta A[\rho]}{\delta \rho(x)} - \int \frac{\partial \rho_N(x')[\rho]}{\partial N} \frac{\delta A[\rho]}{\delta \rho(x')} dx' \ . \tag{A13}$$

Since $\int \frac{\partial \rho_N(x)[\rho]}{\partial N} dx = 1$ due to $\int \rho_N(x)[\rho] dx = N$, it can be seen that the choice

$$u(x) = \frac{\partial \rho_N(x)[\rho]}{\partial N} \tag{A14}$$

gives back Eq.(A1). To justify that $\rho_N[\rho]$ leads to the same second-order constrained derivative as the one given by Eq.(A5), with the choice Eq.(A14), differentiate Eq.(A12) (fully),

$$\frac{\delta^2 \rho_N(x'')[\rho]}{\delta\rho(x)\delta\rho(x')} + \frac{\partial \delta\rho_N(x'')[\rho]}{\partial N \delta\rho(x)} + \frac{\partial \delta\rho_N(x'')[\rho]}{\partial N \delta\rho(x')} + \frac{\partial^2 \rho_N(x'')[\rho]}{\partial N^2} = 0 \ . \tag{A15}$$

Then use Eq.(A15) in the second derivative of $A[\rho_N[\rho]]$ with respect to $\rho(x)$ (see the second line of Eq.(70)). This will give the same expression as Eq.(A5) with Eqs.(14) and

$$\frac{\delta u(x')}{\delta\rho(x)} = \frac{\partial^2 \rho_N(x')[\rho]}{\partial N^2} + \frac{\partial \delta\rho_N(x')[\rho]}{\partial N \delta\rho(x)} \tag{A16}$$

inserted. The higher-order cases can be similarly verified; thus, we have shown that from Eq.(A2), with an arbitrary $u(x)[\rho]$ integrating to one, any proper constrained derivative of any order can be derived.

As an example, the mapping $n_N[n]$ proposed in [26] to give an extension of the DFT functional $F_N[n]$ from the density domain of $\int n(\vec{r})d\vec{r} = N$ may be mentioned. In [26], $n_N[n]$ is defined through an ensemble generalization of $v[n]$ for noninteger $N$'s, as $n_N[n] = n(\vec{r})[N, v[n]]$. [$n(\vec{r})[N, v]$ is the ground-state $N$-electron density in the potential $v(\vec{r})$ – assuming non-degeneracy.] For this $n_N[n]$, $u(\vec{r}) = \dfrac{\partial n(\vec{r})[N, v]}{\partial N}$ is the Fukui function [20], an important chemical reactivity index. From Eq.(A12), we have

$$\int \frac{\delta n(\vec{r}')[N,v]}{\delta v(\vec{r}'')} \frac{\delta v(\vec{r}'')[n]}{\delta n(\vec{r})} d\vec{r}'' = \delta(\vec{r}' - \vec{r}) - \frac{\partial n(\vec{r}')[N,v]}{\partial N} \ , \tag{A17}$$

instead of a simple $\delta(\vec{r}' - \vec{r})$ on the right side. Note that if we put an $N$-conserving constraint on the differentiation, $\dfrac{\partial n(\vec{r}')[N,v]}{\partial N}$ will vanish in Eq.(A17), but at the same time, an $u(\vec{r}')$ will appear in its place (or *if*, for the given application, the use of an ambiguous restricted differentiation suffices, an arbitrary function $g(\vec{r}')$ – which, in the end, can be chosen to be zero).

Finally, we give the generalization of Eq.(A12),

$$\frac{\delta \rho_C(x')[\rho]}{\delta \rho(x)} + \frac{\partial \rho_C(x')[\rho]}{\partial C}\frac{\delta C[\rho]}{\delta \rho(x)} = \delta(x'-x) \ . \tag{A18}$$

That is, $u(x)$ of

$$\frac{\delta A[\rho]}{\delta_C \rho(x)} = \frac{\delta A[\rho]}{\delta \rho(x)} - \frac{\delta C[\rho]}{\delta \rho(x)} \int \left( u(x') \bigg/ \frac{\delta C[\rho]}{\delta \rho(x')} \right) \frac{\delta A[\rho]}{\delta \rho(x')} dx' \tag{A19}$$

[14] can be given as

$$u(x) = \frac{\delta C[\rho]}{\delta \rho(x)}\frac{\partial \rho_C(x)[\rho]}{\partial C} \ . \tag{A20}$$

Eq.(A20) indeed integrates to one, as can be seen by an application of the chain rule of differentiation. In the case of a constraint (5), Eq.(A2) can be generalized, too, yielding

$$\rho_K[\rho] = f^{-1}\left(f(\rho(x)) - u(x)\left(\int f(\rho(x'))dx' - K\right)\right) . \tag{A21}$$

The corresponding mapping for the complex constraint of Eqs.(21) and (22) is given by

$$h_K[h,\phi] = h(x) - u_1(x)\left(\int h(x')dx' - N\right) \tag{A22a}$$

and

$$\phi_K[h,\phi] = \phi(x) - \frac{u_2(x)}{h_K[h,\phi](x)}\left(\int \phi(x')h_K[h,\phi](x')dx' - B\right) . \tag{A22b}$$

**Acknowledgements:** This work was supported by the grant D048675 from OTKA.

[3] T. Gál, Phys. Rev. A **63**, 022506 (2001)

   T. Gál, J. Phys. A **35**, 5899 (2002).

[4] N. Clarke, Eur. Phys. J. E **14**, 207 (2004)

[5] N. Clarke, Macromolecules **38**, 6775 (2005)

[6] R. Yerushalmi-Rozen, T. Kerle, J. Klein, Science **285**, 1254 (1999)

   H. Wang, R. J. Composto, J. Chem. Phys. **113**, 10386 (2000)

   H. J. Chung, R. J. Composto, Phys. Rev. Lett. **92**, 185704 (2004)

   Y. Liao, Z. Su, Z. Sun, T. Shi, L. An, Macromol. Rapid Commun. **27**, 351 (2006)

[7] H. P. Fischer, W. Dieterich, Phys. Rev. E **56**, 6909 (1997)

[8] A. Sharma, J. Mittal, Phys. Rev. Lett. **89**, 186101 (2002)

[9] S. N. Punnathanam, D. S. Corti, J. Chem. Phys. **119**, 10224 (2003)

[10] A. Pototsky, M. Bestehorn, D. Merkt, U. Thiele, Phys. Rev. E **70**, 025201 (2004)

[11] See, e.g., D. J. Evans and G. P. Morriss, *Statistical Mechanics of Nonequilibrium Liquids* (Academic, London, 1990)

[12] M. J. Uline, D. S. Corti, J. Chem. Phys. **129**, 234507 (2008)

[13] P. Blanchard, E. Brüning, *Variational Methods in Mathematical Physics: A Unified Approach* (Springer-Verlag, Berlin, 1992)

[14] T. Gál, J. Math. Chem. **42**, 661 (2007)   [arXiv:math-ph/0603027]

[15] T. Gál, Phys. Lett. A **355**, 148 (2006)  (Note that in Eqs.(25) and (26), an irrelevant factor $\frac{1}{2}$ is missing before the second derivative.)

[16] M. R. Hestenes, *Calculus of Variations and Optimal Control Theory* (Wiley, New York, 1966)

[17] H. T. Davis, *Statistical Mechanics of Phases, Interfaces and Thin Films* (Wiley, New York, 1995)

[18] J. H. Maddocks, Arch. Rat. Mech. Anal. **99**, 301 (1987)
39